\documentclass[12 pt]{article}
 \usepackage{amsmath, amsthm, amssymb, bigints, bm, setspace}
 \usepackage{color}
 \usepackage[margin=1.0in]{geometry}
  
\newcommand{\bfb} {\mbox{\boldmath ${\bm \beta}$}}

\newcommand{\bfg} {\mbox{\boldmath $\gamma$}}
\newcommand{\bfk} {\mbox{\boldmath $\kappa$}}
\newcommand{\bfT} {\mbox{\boldmath $\Theta$}}

\newcommand{\bfl} {\mbox{\boldmath $\lambda$}}

\newcommand{\bfe} {\mbox{\boldmath $\epsilon$}}
\newcommand{\bfta} {\mbox{\boldmath $\tau$}}

\newif\ifpdf
	\ifx\pdfoutput\undefined
	\pdffalse 
	\else
	\pdfoutput=1 
	\pdftrue
	\fi

	\ifpdf
	\usepackage[pdftex]{graphicx}
	\else
	\usepackage{graphicx}
	\fi

\begin{document}

\ifpdf
	\DeclareGraphicsExtensions{.pdf, .png, .jpg, .jpeg, .tif}
	\else
	\DeclareGraphicsExtensions{.eps, .jpg}
	\fi

\begin{center}
\begin{spacing}{1}
{\Large {\bf Bayesian  Variable Selection with Structure Learning: Applications in Integrative Genomics \\}} 
\bigskip
Suprateek Kundu$^3$, Yichen Cheng$^{*1}$, Minsuk Shin$^{*1}$, Ganiraju Manyam$^2$, Bani K. Mallick$^1$, and Veerabhadran Baladandayuthapani$^2$.
\end{spacing}
\end{center}

{\noindent {\bf \quad Abstract.}}  
Significant advances in biotechnology have allowed for simultaneous  measurement of molecular data points across multiple genomic and transcriptomic levels from a single tumor/cancer sample. This has motivated systematic approaches to integrate multi-dimensional structured datasets since cancer development and progression is driven by numerous co-ordinated molecular alterations and the interactions between them. We propose a novel two-step Bayesian approach that combines a variable selection framework with integrative structure learning between multiple sources of data. The structure learning in the first step is accomplished through novel joint graphical models for heterogeneous (mixed scale) data allowing for flexible incorporation of prior knowledge. This structure learning subsequently informs the variable selection in the second step to identify groups of molecular features within and across platforms associated with outcomes of cancer progression. The variable selection strategy adjusts for collinearity and multiplicity, and also has theoretical justifications. We evaluate our methods through simulations and apply them to a motivating genomic (DNA copy number and methylation) and transcriptomic (mRNA expression) data for assessing important markers associated with Glioblastoma progression. 
\bigskip

\begin{spacing}{1}
\begin{footnotesize}
\end{footnotesize}
\bigskip

\end{spacing}

\newpage

\begin{center}{\bf 1. INTRODUCTION} \end{center}


The last decade has seen a proliferation of genomic data, aided partly by the rapid evolution and declining costs of modern technologies,  producing high-throughput multi-dimensional data. It is now technologically and economically feasible  to collect diverse data on matched patient/tumor samples at a detailed molecular resolution  across multiple modalities such as genomics (DNA copy number and methylation), transcriptomics (mRNA/gene expression) and proteomics. Such large scale co-ordinated efforts include worldwide consortiums such as the International Cancer Genome Consortium  (ICGC; icgc.org) and  The Cancer Genome Atlas (TCGA; cancergenome.nih.gov), which have collated data over multiple types of cancer on diverse molecular platforms, to accelerate discovery of molecular markers associated with cancer development and progression.

Initial studies in cancer genomics relying on single platform analyses (mostly gene expression- and protein-based) have discovered multiple candidate ``druggable" targets such as KRAS mutation in colon and lung cancer (Capon, et al., 1983; Shimizu, et al., 1983), BRAF in colorectal, thyroid, and melanoma cancers (Davies, et al., 2002), and PI3K in breast, colon  and ovarian cancers (Samuels, et al., 2004; Bachman et al., 2004; Campbell et al., 2004).  However, it is believed that integrating data across multiple molecular platforms  has the potential to discover more co-ordinated changes on a global (unbiased) level (Chin et. al, 2011). Through data integration, we espouse the philosophy that cancer is driven by numerous molecular/genetic alterations and the interactions between them, with each type of alteration likely to provide a unique but complementary view of cancer progression. This offers a more holistic view of the genomic landscape of cancer, with increased power and lower false discovery rates in detecting important biomarkers (Tyekucheva et al., 2011; Wang et. al, 2013), and translating to substantially improved understanding, clinical management and treatment (Hamid et. al, 2009). Integrating data across diverse platforms has sound biological justifications due to interplay between diverse genomic features. Between platforms, attributes at the DNA level such as  methylation and copy number variation can affect mRNA expression, which in turn is known to influence clinical outcomes such as cancer progression times and stage/pathology of the tumors. Within platform interactions arise from pathway-based dependencies as well as dependencies based on chromosomal/genomic location (an example being copy number data).

Statistically, the concept of data integration can be very broad depending on the scientific question of interest such as prediction, classification, variable selection, or clustering (or combination of these) -- and is an area of active investigation. Lanckriet et al. (2004) propose a two stage approach, first computing a kernel representation for data in each platform and subsequently combining kernels across platforms in a classification model. Shen et. al (2013) proposed a classification model ``iCluster", which uses a joint latent variable model to integrate data across multiple platforms. Tyekucheva et al. (2011) propose a logistic regression model regressing a clinical outcome on covariates across multiple platforms. Recently Wang et al. (2013), Jennings et al (2013) proposed integrative Bayesian analysis of genomics data (iBAG, in short), which models biological relationships between genomic features from multiple platforms, and subsequently uses the estimated relationships to relate the platforms to a clinical outcome. Lock et al. (2013) propose an additive decomposition of variation approach consisting of low-rank approximations capturing joint variation across and within platforms, while using orthogonality constraints to ensure that patterns within and across platforms are unrelated. 

Except for iBAG, the above approaches are not equipped to exploit the information garnered from data integration in relating the platforms to the clinical outcome -- which is usually the goal of translational research in finding markers of cancer progression. Unfortunately, iBAG assumes independence between genes in discovering mechanistic relationships between platforms at a gene-centric level. It is well-established that genes that promote or inhibit tumor development function in co-ordination within modules, such as functional or cell signaling pathways (Boehm and Hahn, 2011).  In this article, we propose an approach that allows for a broad network of interactions within and between platforms (and genes) through a graphical model approach, and subsequently ensures that associated sub-groups of features under the estimated graph influence the outcome in a coordinated manner via a structured variable selection step. Figure 1 provides a conceptual schematic of our approach. 
\begin{figure}
\centering
		\mbox{\includegraphics[height=5in, width= 1 \textwidth]{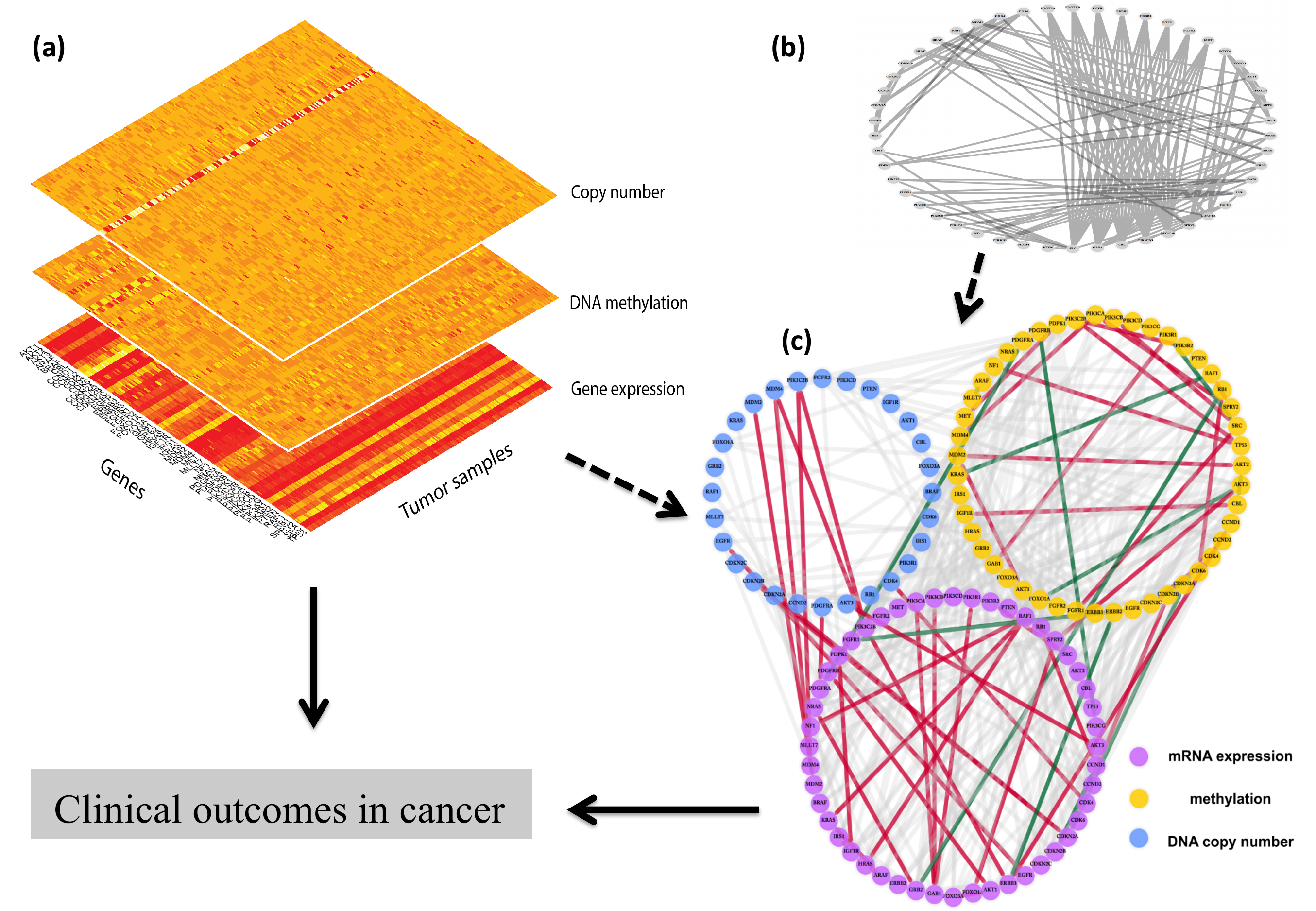}} 		
		\caption{\small{A schematic diagram of our integrative modeling approach. Panel (a) shows the heatmaps of the genes by sample matrix constructed from data for three platforms; panel (b) depicts the prior graph constructed using previous
studies; while panel (c) is the estimated graph of the genes within and across the platforms. The dashed arrows determine graphical structure and the solid arrows represent the regression model incorporating graphical dependencies. Red and green lines in panel (c) represent high negative and positive partial correlations under the estimated graph, while all other edges with lower absolute partial correlations are depicted with watermark lines.}}
\end{figure} 

Our methods are motivated by a TCGA based study focusing on discovery of important molecular markers for progression times in glioblastoma patients by integrating diverse platform-specific features on different scales: discrete (copy number variation) and continuous (DNA methylation and mRNA expression), measured on the same set of genes. In addition, there exists substantial prior knowledge on pathway (graphical) interactions between these genes (e.g. from databases such as KEGG, Gene Ontology or from other prior studies), the incorporation of which can result in improved estimation and refined biological interpretations. Thus integrating information across these platforms requires us to develop a graphical modeling approach which estimates the graph for mixed data (from different scales), while being able to incorporate prior graphical knowledge. Different platforms might be potentially informative about different sets of within platform gene--gene interactions which are not evident based on single platform analysis, while clusters of features within and across platforms are expected to work together to drive the cancer mechanism, thus rendering the usual variable selection approaches assuming independence between covariates inadequate. 

There is an increasing variable selection literature for structured covariates lying on a known graph, which mirrors the growing recognition that incorporation of supplementary covariate information in the analysis of genomic data can be instrumental in improved inferences (Pan et al., 2010). Such developments have been aided by a proliferation of genomic databases storing pathway and gene-gene interaction information (Stingo et. al, 2011). 
Some examples of Bayesian regression and variable selection approaches for graph structured covariates include Li and Zhang (2010), Stingo et al. (2011), and Rockova and Lesaffre (2013). However, the focus of the above approaches has been on single platform regression with covariates lying on a {\it known} graph specified {\it a priori}, which is very different from the data integration component in the present work, whereby we seek to achieve the dual goal of learning the graph for mixed covariates while incorporating prior knowledge, and subsequently using such structure learning to relate the covariates to the outcome. We note that naively using a fixed known graph may not be practical in genomic studies, where the network of gene--gene interactions is likely to vary over different biological or experimental conditions. In such cases, mis-specification of prior information can result in inferior quality inferences.


We propose a principled two stage approach which estimates the graph in the first step, and uses this point estimate to inform the subsequent variable selection step. Concisely stated, the major novelties presented in our article are: (i) estimating graphical models for mixed data, while incorporating prior graphical knowledge and controlling the degree of confidence one wishes to place on the prior graph through a belief parameter. The graphical modeling approach for mixed data models ordered categorical variables by rounding continuous latent variables (Albert and Chib, 1993; Kottas, 2005; Jara et. al, 2007; Canale and Dunson, 2012), and specifies a graphical model jointly on the observed continuous covariates and the latent continuous variables; (ii) using the estimated graph to define (potentially overlapping) groups of molecular  features within and across platforms, which work together in driving the outcome; and (iii) developing a structured variable selection approach, which is designed to control collinearity by simultaneously including all variables belonging to each aforementioned group, while having sufficient control over false positives by specifying appropriate multiplicity adjusted priors. The proposed approach generalizes the afore-mentioned approaches for graph-structured covariates assuming a fixed known graph when the belief parameter is large ($\approx\infty$), and corresponds to an unsupervised approach with complete lack of prior graphical knowledge when the belief is small ($\approx 0$). In addition, to our knowledge, ours is the first Bayesian graphical model approach to address mixed data.

Section 2 describes the methodology, section 3 outlines the posterior computation scheme, section 4 provides theoretical justifications for our approach, section 5 lays out simulation studies, section 6 applies our approach to genomic data for individuals with glioblastma multiforme. A conclusion section summarizing the main results is provided and the Appendix contains proof of the Theorems and posterior computation steps.

\vskip 12pt

\begin{center}{\bf 2. METHODOLOGY} \end{center}

{\noindent \bf Notations:} In this article, we focus on a univariate continuous response, $y \in \Re$, to be regressed on a $p$-dimensional vector of mixed covariates ${\bf x} = [\bf x_{1},\ldots, \bf x_{D^*}]$ having an underlying graphical structure which is to be estimated, while incorporating prior graphical knowledge $G_0$. Here ${\bf x}_j$ is the covariate vector corresponding to the j-th platform which has $p_j$ features, $j=1,\ldots,D^*$, so that $\sum_{j=1}^{D^*} p_j=p$. The discrete covariates (e.g. copy number) in the mixed vector ${\bf x}$ are allowed to be binary or ordered categorical. 
 
Let us denote the $n\times 1$ vector of responses as $Y^n$ and the $n\times p$ dimensional covariate matrix as $X=[ X_{1},\ldots, X_{D^*}]$. We construct a joint probability model employing a two stage strategy, and based on the factorization $P(Y^n,X)=P(X)P(Y^n|X)$. In the first structure learning stage,  we use the posterior \begin{small}$P(\Omega|X,G_0)\propto P(X|\Omega)P(\Omega|G_0)$\end{small} to estimate the $p\times p$ precision (inverse covariance) matrix $\Omega$ for the covariates, under a continuous shrinkage prior $P(\Omega|G_0)$ incorporating prior graph knowledge $G_0$. The precision matrix is used to obtain  estimates of the graph $\hat{G}$ using a post--MCMC step as described in the sequel. In the second step, called structured variable selection, $\hat{G}$ is used to inform the regression model through the prior on the model space, i.e. we specify the likelihood  \begin{small}$P(Y^n|X,M_{\bfg},\bfT^{Y|X})P(M_{\bfg}|\hat{G},\bfT^{Y|X})P(\bfT^{Y|X}|\hat{G})$\end{small}, where $M_{\bfg}$ denotes a model in the model space $\Gamma$ and $\bfT^{Y|X}$ is the list of regression parameters. The goal of the second step is to propose a variable selection model designed to control collinearity by incorporating the information represented by the graph structure on covariates, while having sufficient control over multiplicity through multiplicity adjusted priors. We discuss these steps in detail below. \\

{\noindent {\bf 2.1 Integrative Structure Learning using Mixed Graphical Models}}\\
Recalling that we are dealing with mixed covariates ${\bf x} = [\bf x_{1},\ldots, x_{D^*}]$ across diverse platforms, we propose a graphical model approach for mixed data designed to integrate and find inherent structures within and across multiple platforms. The graphical modeling approach involving the vertex set $V=\left\{1,\ldots,p\right\}$ and edge set $E$, serves the following two purposes: (1) model dependence between features within and across platforms and genes -- in our application, measurements for different platforms are available for the same set of genes, so that the joint modeling across platforms allows for both cis-acting (localized to a gene) and trans-acting (across gene locations)  ; and subsequently (2) use the graph to detect subgroups of features within and across platforms, as well as gene locations, which define functional modules that are related and potentially work together to drive cancer progression. Such modules correspond to cliques, which are defined as a subgroup of $V$ such that each node in this subgroup is connected to every other node in the subgroup. 



{\noindent \bf The mixed data formulation:} Without loss of generality, let ${\bf x}_i = [\bf x_{i1},\ldots, x_{iD^*}]=({\bf x}^C_i, {\bf x}^O_i)$ denote the covariate vector for the i-th subject, with the superscripts $C,O$ denoting continuous and ordinal (and/or binary) covariates respectively. Let $z^O$ denote the generic notation for the latent continuous variable corresponding to ordinal predictor $x^O$, and consider the following graphical model for mixed covariates 
\begin{small}
\begin{eqnarray}
x^O_{ij} &=& l, \mbox{ if } D_{l-1}<= z^O_{ij} < D_{l}, \quad -\infty = D_0<D_1<\ldots<D_{M_o} = \infty, \quad j=1,\ldots,p_O,  \nonumber \\
{\bf z}^O_i &=& (z^O_{i1},\ldots,z^O_{ip_O}), \quad ({\bf x}^C_i,{\bf z}^O_i) \sim N_{[D]}(0,\Omega^{-1}), \quad \Omega\sim \pi(\Omega), \quad i=1,\ldots,n, \label{eq:covmodel}
\end{eqnarray}
\end{small}
where $N_{[D]}$ denotes a Gaussian distribution with truncated domains for ${\bf z}^O$, $M_o$ is the number of ordinal levels, $p_O$ is the number of ordinal covariates, and $\Omega\sim\pi(\Omega)$ corresponds to a continuous shrinkage prior on the $p\times p$ precision matrix to be described in the sequel. To keep things simple, the prior on the precision does not incorporate prior graphical knowledge $G_0$ at this stage, which is to be introduced in the next section. Instead of specifying shrinkage priors, one can alternatively choose a discrete mixture specification (Dawid and Lauritzen, 1993; Talluri et al., 2013) in (\ref{eq:covmodel}) as: \begin{small} $({\bf x}^C_i,{\bf z}^O_i) \sim N_{[D]}(0,\Omega^{-1}_G), \quad \Omega_G\sim \pi(\Omega|G),\quad G\sim\pi(G), \quad i=1,\ldots,n.$ \end{small} Such discrete mixture approaches would place exact zeros for off-diagonals with absent edges corresponding to $G$, while maintaining positive definiteness of $\Omega_G$ and simultaneously specifying a prior $G\sim \pi(G)$. However, we adopt a continuous shrinkage approach, as it provides us an avenue for incorporating prior knowledge in a seamless manner as well as achieving sparsity. Under the generic continuous shrinkage specification (\ref{eq:covmodel}), the posterior is given by
\begin{small}
\begin{eqnarray*}
P(\Omega,{\bf z}^O_1,\ldots,{\bf z}^O_n|X)\propto \pi(\Omega)\prod_{i=1}^{n}\left\{\prod_{j=1}^{p_O}\sum_{l=1}^{M_o} 1(x^O_{ij}=l)1(D_{l-1}\le z^O_{ij}<D_l)\right\}N({\bf x}^C_i,{\bf z}^O_i;0;\Omega^{-1}),
\end{eqnarray*}
\end{small}
from which MCMC samples arise. Subsequently a post-MCMC step can be implemented in order to obtain the graph estimate $\hat{G}$ by thresholding absolute partial correlations corresponding to the estimated precision matrix $\hat{\Omega}$, as elaborated in Section 3. 

Unfortunately, in spite of a rich literature on Gaussian graphical models, there seems to be limited development of graphical modeling approaches incorporating prior knowledge. The most common approach seems to be incorporating the prior graph via informative priors on edge inclusion probabilities (Baladandayuthapani et. al, 2014), with Mukherjee and Speed (2008) providing an alternative approach. In general, incorporating prior knowledge for all edges through a discrete mixture approach can run in to potential difficulties, since finite runs of the MCMC are not able to update a sizable proportion of the edges even for moderate dimensional graphs. In such a case, these edges will not be updated at all, and will instead correspond to the initial choice of the adjacency matrix (Kundu et al., 2014). This violates our objective of incorporating prior knowledge while learning all possible edges of the graph from the data, with such discrete mixture approaches potentially resulting in graphical estimates which are increasingly sensitive to the initial choice of the adjacency matrix for higher dimensions and finite lengths of MCMC chains.

The above considerations motivate us to use a continuous shrinkage approach for graphical models, which is equipped to incorporate prior graphical knowledge by using a novel procedure involving a belief parameter. Unlike discrete mixture approaches, the continuous shrinkage based approach does not depend on an initial adjacency matrix, and can update all elements of the precision matrix at every iteration, thus utilizing the available prior knowledge on all edges to drive inferences on the graph.  \\


{\noindent{\bf 2.2 Incorporating Prior Graph Information}}\\
We now describe the graphical prior incorporating prior knowledge. Let $G_0$ be the prior graph having vertex set $V=\left\{1,\ldots,p\right\}$ and edge set $E_0$, with the corresponding adjacency matrix $A_0=(a_{0,ij})$, where $a_{0,ij}=1$ if edge $(i,j)$ is present and 0 otherwise. Throughout this article, we consider undirected graphs so that $a_{0,ij}=a_{0,ji}$ for all $(i,j)$. In order to incorporate prior knowledge $G_0$, we develop the following novel generalization of the Bayesian graphical lasso (Wang, 2012)
\begin{small}
\begin{eqnarray}
\pi(\Omega|\bfta) &\propto & \prod_{i<j}\frac{1}{\sqrt{2\pi\tau_{ij}}}\exp(\frac{-\omega^2_{ij}}{2\tau_{ij}})\prod_{i=1}^p \frac{\lambda_{ii}}{2}\exp(-\frac{\lambda_{ii}}{2}\omega_{ii}) 1_{\Omega\in M^{+}},  \nonumber \\
\pi(\bfta|\bfl) &\propto& \prod_{i<j} \frac{\lambda_{ij}^2}{2}\exp\bigg(-\frac{\lambda_{ij}^2}{2}\tau_{ij} \bigg), \quad \lambda_{ij}|G_0 \sim (1-p_{ij})Ga (\kappa_{ij} + a_\lambda, b_\lambda) + p_{ij} Ga(a_\lambda,b_\lambda),\nonumber \\
 p_{ij} &\sim& Be(a_{0,ij}\kappa_{ij} + a_p, (1-a_{0,ij})\kappa_{ij} + b_p), \label{eq:priorgraph}
\end{eqnarray}
\end{small}
where $M^+$ is the set of positive definite matrices, $\bfl,\bfta,$ are vectorized parameters with dimension $p(p+1)/2$, and $\kappa_{ij}$ is the belief parameter for edge (i,j) ($\kappa_{ij}=\kappa_{ji},i\ne j$, under an undirected graph). In (\ref{eq:priorgraph}), $\pi(\Omega|\bfta)$ and $\pi(\bfta|\bfl)$ specifiy a double exponential prior $\omega_{ij}\sim\mbox{DE}(\lambda_{ij}),i\ne j$ and an exponential prior on the diagonals $\omega_{ii}\sim\mbox{Exp}(\lambda_{ii})$. The novelty of our approach involves incorporating prior knowledge $G_0$ through the shrinkage parameter $\bfl$ via the belief parameters $\bfk$. Specification of different beliefs for different edges can be useful when subject knowledge or historical information points strongly towards presence or absence of certain edges (gene--gene interactions in our application), but is ambiguous about others, as often happens in genomic studies. 

{\noindent \underline{Role of belief parameter}:} Through the use of a belief parameter, we can control the degree of confidence we place on the available prior graph information. This is a useful feature in enabling investigators to be flexible i.e. either skeptical or fairly confident about the prior knowledge, thus providing them with a range of alternatives in situations where there is reason to suspect mis-specification. To understand the role of the belief parameter in prior specification, observe that 
\begin{small}
\begin{eqnarray}
E(p_{ij})= (a_{0,ij}\kappa_{ij} + a_p)/(\kappa_{ij} + a_p + b_p), \quad E(\lambda_{ij}) = (1-p_{ij})(\kappa_{ij} + a_\lambda)/b_\lambda + p_{ij}a_\lambda/ b_\lambda, \label{eq:calibrate}
\end{eqnarray}
\end{small}
where $i\ne j=1,\ldots,p$. When $a_{0,ij}=1$ (i.e. $G_0$ suggests presence of an edge), $E(p_{ij})\approx 1$ for large values of $\kappa_{ij}$ with $\kappa_{ij}>>b_p$. In extreme case when $\kappa_{ij}\to\infty$, we have $p_{ij}\approx 1$ which implies $E(\lambda_{ij}) \approx a_\lambda/ b_\lambda \approx 0$ for appropriate values of $a_\lambda,b_\lambda$. Again for a large $\kappa_{ij}$ and for $a_{0,ij}=0$ (i.e. $G_0$ suggests absence of an edge), we have $E(p_{ij})\approx 0$. In the extreme case when $\kappa_{ij}\to\infty$, we have $p_{ij}\approx 0$ which implies $E(\lambda_{ij})\approx (\kappa_{ij} + a_\lambda)/b_\lambda\to\infty$. In summary, a large value of $\kappa_{ij}$ encourages the presence or absence of an edge through a small or large value of $\lambda_{ij}$, according to $a_{0,ij}=$1 or 0 in the prior graph. In the limiting case as $\kappa\to\infty$, the realizations under formulation (\ref{eq:priorgraph}) will become increasingly degenerate at the prior graph $G_0$. On the other hand when we are skeptical of the prior graph knowledge, one can simply choose $\kappa\approx 0$ which reduces formulation (\ref{eq:priorgraph}) to \begin{small}$\lambda_{ij} \sim Ga(a_\lambda,b_\lambda)$\end{small}, thus collapsing to the set-up with no prior knowledge. These two scenarios correspond to two extremes of regression for graph structured covariates in supervised and unsupervised settings, assuming a fixed known graph and a complete lack of prior knowledge respectively.


{\noindent \underline{Calibration of belief parameter}:} In practice we can use $E(p_{ij})$ in (\ref{eq:calibrate}) to quantify the degree of confidence on the presence of an edge specified by prior knowledge. On rearranging the terms in (\ref{eq:calibrate}) one has 
\begin{small}
\begin{eqnarray*}
\kappa_{ij} = b_p/(1-E(p_{ij})) - (a_p+b_p), \mbox{ when } a_{0,ij}=1,  \quad \kappa_{ij}= a_p/E(p_{ij}) - (a_p+b_p), \mbox{ when } a_{0,ij}=0.
\end{eqnarray*}
\end{small}
Thus, if $G_0$ suggests presence of the edge $(i,j)$, and we are 50\% sure of the correctness of such prior knowledge, we can substitute $E(p_{ij})=0.5$ in the above expression to obtain $\kappa_{ij}=b_p-a_p$. In practice, one can choose a suitable combination of $\kappa_{ij},a_p,b_p$ to reflect appropriate confidence. Figure 2 presents a plot of the belief parameter versus the degree of confidence for the scenarios when the prior graph suggests presence ($a_{0,ij}=1$) and absence ($a_{0,ij}=0$) of the (i,j)-th edge. From the plot, one can categorize different regions of confidence: for $a_{0,ij}=1$, low confidence corresponds to \begin{small}$E(p_{ij})<0.5(-1\le \kappa_{ij}< 0)$\end{small}, moderate confidence corresponds to \begin{small}$0.5\le E(p_{ij})\le 0.7(0\le\kappa_{ij}\le 1.34)$\end{small}, and high confidence corresponds to \begin{small}$E(p_{ij})>0.7(\kappa_{ij}>1.34)$\end{small}. Similar regions can also be constructed for $a_{0,ij}=0$. 

\begin{figure}
\centering
		\mbox{\includegraphics[height=2.5in, width= 1 \textwidth]{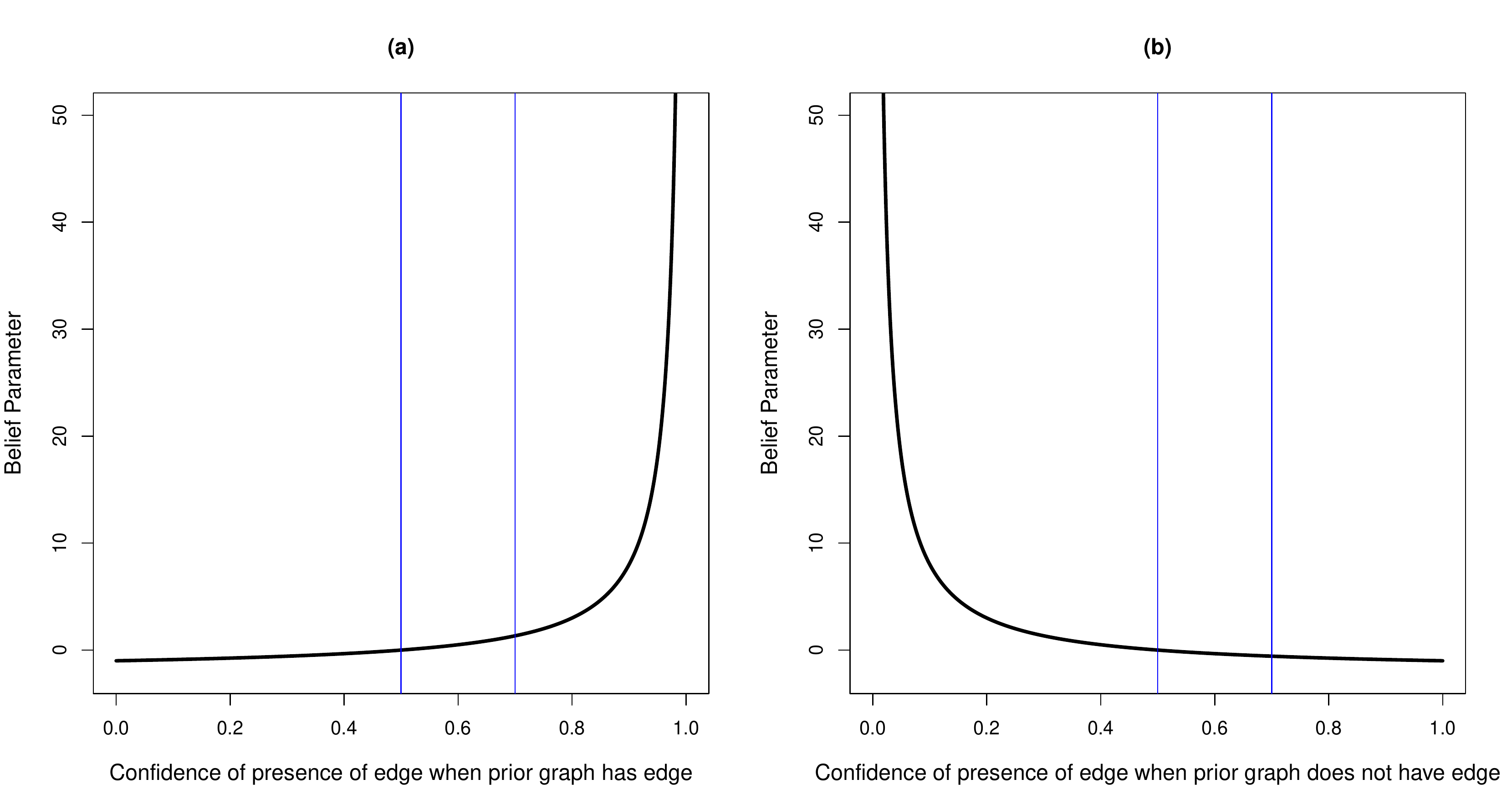}} 		
		\caption{\small{Calibration of belief parameter. The plot shows the range of the the belief parameter (y-axis)
versus degree of confidence (x-axis). The three bins demarcated by vertical blue lines depict regions of low ($-1\le \kappa< 0$), moderate ($0\le \kappa<1.34$) and high confidence ($\kappa>1.34$) from left to right.}}
\end{figure} 

{\noindent{\bf 2.3. Regression and Structured Variable Selection}} \\
Through the above steps we obtain an estimate of the graphical structure on the covariates represented as $\hat{G}$. In the second step, called structured variable selection, we incorporate the structural knowledge represented by the estimated graph in regressing the outcome of interest on covariates. Although we consider continuous outcomes, it is straightforward to extend our approach to binary or ordinal outcomes via thresholding the latent continuous variables (see citations in Introduction). We consider the following linear regression model 
\begin{small} 
\begin{eqnarray}
 Y^n &=& \alpha{\bf 1}_n + {\bf X}_{{\bfg}}{\bfb}_{\bfg} + \bfe,\quad \epsilon_i \sim N(0,\eta^{-1}), \quad \pi(\alpha,\eta)\propto 1/\eta,    \nonumber \\
\beta_\gamma &\sim& N\left( 0, g\eta^{-1}({\bf {\bf X}}_{{\bfg}}' {\bf X}_{{\bfg}})^{-1} \right), \quad \pi(g) = \frac{a-2}{2}(1+g)^{-\frac{a}{2}}, \quad \bfg \sim \pi(\bfg|\hat{G}), \quad i=1,\ldots,n. \label{eq:base} 
\end{eqnarray}
\end{small}
Here ${\bm \gamma} = \{ \gamma_j, j=1,\ldots, p\} \in \Gamma$ (the model space) is the vector of variable inclusion indicators, with $\gamma_j =1$ if the $j$th candidate predictor is included in the model and $\gamma_j=0$ otherwise, ${\bm\beta}_{\bm\gamma}$ is the $p_{\bfg}\times 1$ vector of the regression coefficients with $p_{\bfg} = \sum_{j=1}^p \gamma_j$ being the size of model $\bfg$, ${\bf X}_{\bfg}$ is the $n \times p_{\bfg}$ covariate matrix (excluding an intercept) containing the predictors in model ${\bfg}$ and having the i-th row as ${\bf x}_{\bfg,i}$, and $\alpha,\eta$ are the intercept and the residual precision respectively. We address uncertainty in subset selection through $\pi(\bfg|\hat{G})$ depending on the estimated graph structure on the covariates, while mixtures of $g$-prior (Liang et al., 2008) on $\bfb_{\bfg}$ characterizes the prior knowledge of the size of the coefficients for the selected predictors. The tuning parameter $g$ is assigned the hyper-$g$ prior with $a=4$.  

The prior on the model space $\bfg\sim\pi(\bfg|\hat{G})$ is defined using clique indicators. Let $C_1,\ldots,C_q$, denote the cliques identified by the estimated graph $\hat{G}$. The cliques are indicative of (potentially overlapping) groups of associated genetic features within and across platforms and gene locations. The variable selection approach espouses the philosophy that features in each clique are indicative of functional modules which work in coordination to drive the outcome. To this end, clique inclusion indicators are designed to include all features in a clique simultaneously if that clique is significantly associated with the outcome. Denote inclusion indicators for the k-th clique as $\gamma_{C_k}$, $k=1,\ldots,q$.  We define the prior on the model space through clique inclusion probabilities as follows
\begin{eqnarray}
P(\gamma_{C_k}=1|\hat{G}) = \pi, \quad \pi \sim Be(a_{\pi},b_{\pi}),  \label{eq:BVSprior}
\end{eqnarray}
where $\pi$ controls the sparsity of the model through the Beta hyperprior, with the resulting formulation achieving multiplicity control over the selection of cliques (Scott and Berger, 2010). The above approach is designed to protect against collinearity by including a particular feature if any one of the cliques to which it belongs is significant. This implies $P(\gamma_j=1)= P(\cup_{k\in \mathcal{S}^j_{C}}\mbox{ } \{\gamma_{C_k}=1\})$, where $\mathcal{S}^j_{C}$ is the set of all cliques containing the j-th node. We call the resulting approach in (\ref{eq:base})-(\ref{eq:BVSprior})  {\it Bayesian variable selection with structure learning} (BVS-SL), a schematic representation of which is presented in Figure 3.

\begin{figure}
\centering
		\mbox{\includegraphics[height=3.5in, width= 1 \textwidth]{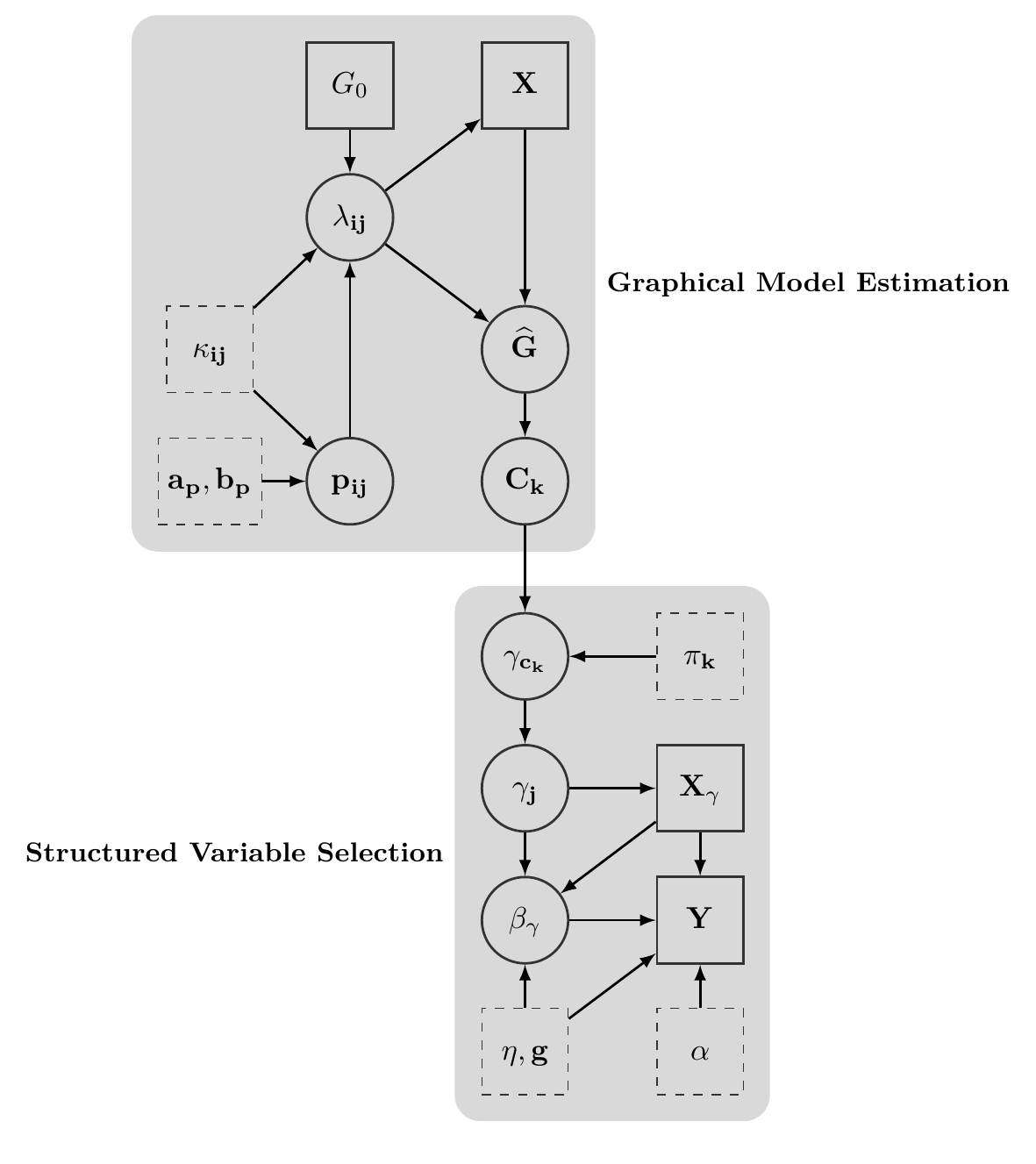}} 		
		\caption{\small{Directed acyclic graph of proposed model. The indices run : $i\ne j=1,\ldots,p,k=1,\ldots,q,$ with $q$ being number of cliques in the estimated graph. Solid boxes, circles, dashed boxes, represent observed data,  model parameters, and model hyperparameters respectively.}}
\end{figure} 

In practice, one can specify different inclusion probabilities $\pi_k,k=1,\ldots,q,$ in (\ref{eq:BVSprior}) to account for the clique size, so as to encourage or discourage inclusion of large cliques. However our approach is designed to produce sparse graphical estimates, so that the clique sizes are expected be small and would not exhibit high variability, with such sparsity assumptions on the graph being pervasive in statistical literature. The simultaneous control of collinearity for features within cliques and multiplicity control over different cliques is designed to attain a desirable balance between detecting true positives and true negatives, a claim which is supported by our simulation studies.

By computing the marginal inclusion probabilities of cliques, we are in a position to detect such subgroups which significantly affect outcome of interest. In addition, individual significant covariates can also be detected by computing marginal inclusion probabilities as described previously. In the scenario when all the cliques are disjoint with $q<p$, the formulation reduces to a clustering approach with all features within a certain cluster either simultaneously included or excluded, while allowing the within cluster fixed effects to be numerically different. In the limiting case, when each of these disjoint cliques has high positive pair-wise correlations tending to 1, our approach reduces to the traditional clustering approach with the within cluster fixed effects becoming numerically similar. In the special case when $q=p$, the above formulation reduces to the usual stochastic search variable selection (SSVS) approach. Finally, we demonstrate that  the variable selection approach has appealing theoretical justifications under suitable assumptions as highlighted in Section 4.

\vskip 12pt

\begin{center} {\bf 3. Posterior Computation } \end{center}

The proposed approach contains two distinct sets of posterior computation, one for the graphical model estimation part, and another for the structured variable selection approach. The graphical model estimation for mixed covariates proceeds via sampling the latent underlying continuous variables corresponding to the ordered discrete covariates, followed by drawing the joint precision matrix of $({\bf x}^C,{\bf z}^O)$ under formulation (\ref{eq:priorgraph}). We adapt the procedure in Johnson and Albert (2001) to the case of dependent covariates, for posterior updates of the latent continuous covariates under the following posterior distributions
\begin{eqnarray*}
z^O_{ij}|x^O_{ij}=l,D_{l-1,j},D_{l,j} \sim N_{(D_{l-1,j},D_{l,j}]}(z^O_{j}|{\bf x}_i^C,{\bf z}_i^O(-j)), \quad D_{l,j}|z^O_j\sim \mbox{Unif}(z^L_{l,j},z^U_{l,j}),
\end{eqnarray*}
where ${\bf z}_i^O(-j)$ represents the vector of latent underlying variables for the i-th subject and excluding the j-th measurement, $z^L_{l,j} = \max\limits_{i:x^O_{ij}=l}z^O_{ij}$ and $z^U_{l,j} = \min\limits_{i:x^O_{ij}=l+1}z^O_{ij}$ for $l=1,...,M-1$. Once the latent variables have been updated at each MCMC iteration, we sample $\Omega,\bfta$ using the method described in Wang (2012), 
while $\lambda_{ij}$ is updated by introducing an auxilary binary variable \begin{small}$\delta_{ij}\sim\mbox{Bernoulli}(p_{ij})$\end{small}, and using the posterior
\begin{small}
\begin{eqnarray*}
\pi(\lambda_{ij}|-)\sim Ga(1+a_\lambda,|\omega_{ij}|+b_\lambda)1(\delta_{ij}=1) +  Ga(1+\kappa_{ij}+a_\lambda,|\omega_{ij}|+b_\lambda)1(\delta_{ij}=0),  
\end{eqnarray*}
\end{small}
where $1(\cdot)$ is an indicator function and $p_{ij}$ is drawn from a Beta posterior. The point estimate of the graph is obtained as a post--MCMC step by thresholding the estimated partial correlations. In particular, the (i,j)-th edge is included if and only if \begin{small}$\hat{\rho}_{ij}/E_{\mathcal{W}}(\rho_{ij}|{\bf X}) > 0.5$, \end{small} where \begin{small}$\hat{\rho}_{ij}$\end{small} is the posterior partial correlation estimate under the continuous shrinkage approach, and \begin{small}$E_{\mathcal{W}}(\rho_{ij}|{\bf X})$\end{small} represents the posterior mean of the partial correlation under the reference distribution $\mathcal{W}=$Wishart$(3,I_p)$.  The posterior computation steps for structured variable selection conditional on $\hat{G}$ are presented in the Appendix A.3.

\vskip 12pt

\begin{center}{ \bf 4. THEORETICAL JUSTIFICATIONS}\end{center}


In this section, we establish variable selection and prediction consistency results for the proposed structured variable selection approach. Model/variable selection consistency implies increasing posterior probability to the true model as the sample size increases, i.e. $P(M_{\bfg_0}|Y^n,{\bf X})\to 1$ as $n\to\infty$, where the true model in our case is given by
\begin{small}
\begin{eqnarray}
M_{\bfg_0}: Y^n = \alpha + {\bf x}_{\bfg_0}\beta_{\bfg_0} + \bfe, \quad \epsilon_i \sim N(0,\eta^{-1}), \quad {\bf x}_i \sim N(0,\Omega^{-1}_0), i=1,\ldots,n. \label{eq:TrueModel}
\end{eqnarray}
\end{small}
For prediction consistency, we would like the predicted outcome to be close to the true expected value with increasing certainty. Given the covariate vector ${\bf x}^*$, and the data $(Y^n,{\bf X})$, the predicted value as defined in Liang et. al (2008) is the optimal point estimator under the squared error loss obtained by Bayesian model averaging 
\begin{small}
\begin{eqnarray*}
\hat{y}^* = \hat{\alpha} + \sum_{\bfg\in \Gamma}{\bf x}^*_{\bfg}\hat{\beta}_{\bfg}P(M_{\bfg}|Y^n,{\bf X})\int_{0}^\infty \frac{g}{1+g}\pi(g|M_{\bfg},Y^n,{\bf X}) dg,
\end{eqnarray*}
\end{small}
where $\hat{\alpha},\hat{\beta}_{\bfg}$ are the ordinary least squares estimates for model $M_{\bfg}$. We call the proposed approach consistent under prediction if $P(\hat{y}^*=E_0(y^*)|{\bf X})\to 1$ as $n\to\infty$ under the true sampling distribution (i.e. when $M_{\bfg_0},\alpha,\bfb_{\bfg_0},\eta$ is known), where $E_0(y^*)=\alpha + {\bf x}_{\bfg_0}\beta_{\bfg_0}$ is the mean in the true model. Denote $M_{N}$ as the null model. We have the following result under regulatory conditions {\em (A1)--(A3)}, which are presented in Appendix A.1, with the proof in Appendix A.2.


{\noindent \bf Theorem 1} {\em Under assumptions {\em (A1)--(A3)} and for the class of models defined by (\ref{eq:base}), model selection consistency and prediction consistency hold under the true model $M_{\bfg_0}(\ne M_{N})$.} \\

\vskip 12pt

\begin{center}{\bf 5. SIMULATIONS} \end{center} 

The goal of our simulation study is to assess the variable selection and prediction performance for the proposed approach under several scenarios with varying dimensions and association structures for the covariates. We implement the proposed approach both without prior knowledge (BVS-SL) and with prior knowledge (prior corrected BVS-SL), where the prior graph is taken to be the true graph $G_0$ used to generate the data. The same value of the belief parameter (50) was used for all edges corresponding to strong confidence; we also examine the effect of varying the belief parameter as explained in the sequel. We compare the proposed approach to stochastic search variable selection (SSVS) (George and McCullough, 1993) assuming independence of predictors, the penalized joint credible regions approach (PenCred) by Bondell and Reich (2012), which takes in to account dependence amongst predictors, as well as frequentist regularization approaches such as Lasso (Tibshirani et al., 2005) and elastic net (Zou and Hastie, 2005), using R packages `lars' and `elasticnet' respectively. To determine the cliques for an estimated graph $\hat{G}$ under our approach, we use the `igraph' package in R. The number of MCMC iterations implemented for Bayesian approaches was 10,000, with a burn in of 3000. The training and test sample sizes were 100 each, and we consider $p=24,40,80$. All results are reported over 20 replicates.

The data was generated from a linear regression model having $p$ covariates out of which nine were ordinal (generated by thresholding the continuous latent variables) and taking values 0-4, and the rest were continuous. The true inclusion status is set to $\gamma^0_j =1,j=1,\ldots,8,23,24,$ with four discrete variables included, and $\gamma^0_j =0$ otherwise. The continuous covariates and the continuous latent variables for discrete covariates were generated using a multivariate Gaussian distribution with covariance $\Sigma_T$. We consider different block-diagonal structures for $\Sigma_T$ (listed hereafter), specifying subgroups of predictors with varying partial correlations. The true graph $G_0$ was obtained by including all edges $(i,j)$ with $|\Sigma_T^{-1}(i,j)|>0.0001$.  \\
{\bf Case I(a):} This case corresponds to high partial correlations with the precision matrix having four sub-blocks and all precision diagonals being 1. The first sub-block ($4\times 4$) has off-diagonals as $0.95$, the second and third sub-blocks ($4\times 4$ each) have precision off-diagonals as $0.7$, and the fourth sub-block ($\overline{p-12}\times \overline{p-12}$) is identity. The true coefficient vector was \begin{small}$(0.3,-0.7,1.1,-0.05,0.1,0.2,-1.2,1.5,0,\ldots,0,1,-1)$.
\end{small}\\
{\bf Case I(b):} This case corresponds to high correlations with $\Sigma_T$ having the same structure as $\Sigma^{-1}_T$ in Case I(a). The coefficients were \begin{small}$(0.3,0.7,1.1,0.05,-0.1,-0.2,-1.2,-1.5,0,\ldots,0,1,-1)$\end{small}. Different signs for coefficients in cases I(a)--(b) ensures same signs for positively pair-wise correlated covariates. 
\\
{\bf Case I(c):} This case corresponds to a block diagonal with two sub-blocks - one having an AR(1) structure for the precision matrix with \begin{small}$\Sigma^{-1}_T(i,j)=0.95^{|i-j|}, i,j=1,\ldots,8,$ \end{small}and the other sub-block being identity. The coefficients were same as those in Case I(a).\\
{\bf Case I(d):} This case corresponds to $\Sigma_T$ having the same structure as $\Sigma_T^{-1}$ in Case I(c). The coefficients were same as those in Case I(b).\\


{\noindent \bf Ordering of Models:} One can obtain an ordered sequence of regression models by varying the cut-off for the marginal inclusion probability under Bayesian approaches and varying the penalty parameter for frequentist approaches. For each point on the ordering, we can obtain specificity = 1- false positive rate (FPR), and sensitivity which can be considered as the power to detect important predictors. To evaluate the ordering of the models, we look at receiver operating characteristic (ROC ) and precision recall characteristic (PRC) curves. For the set of ordered models, ROC curves plot sensitivity versus 1-specificity, while PRC curves instead plot the precision (ratio of true positives to the total number declared as positive) versus the recall or sensitivity. ROC curves present a picture of the trade-off between type I error and power, while PRC curves give a complementary picture, examining the trade-off between the power and false discovery rate (Davis and Goadrich, 2006). 

From the ROC and PRC curves presented in the Supplementary Material, it is clear that the curves for BVS-SL and the prior corrected BVS-SL, essentially always dominate the competing curves for all cases in higher dimensions. The area under ROC and PRC curves presented in Tables 1--4 shows that the BVS-SL and prior corrected BVS-SL approaches essentially always perform better than the competitors for all dimensions, which indicates a superior model selection performance. From the Tables, BVS-SL demonstrates a uniformly higher power when the false discovery rate is controlled at 10\%, which points towards a superior performance in tackling collinearity for a given multiplicity level.

Under the sample size and model dimensions we consider, the prior corrected BVS-SL may or may not perform better than it's uncorrected counterpart. We expect that as $n$ increases, the former will almost always outperform the latter for high values of the belief parameter, under no or minimal mis-specification of the prior knowledge. Sensitivity to prior knowledge is discussed in more detail in the sequel. 

{\noindent \bf Out of Sample Prediction:} In addition to looking at the ordered sequence, it is certainly of interest to examine the predictive performance under each approach, as well as to assess the point estimate under the optimal model. The point estimate is selected using the Bayesian information criterion under BVS-SL, PenCred, Lasso, and elastic net, while the median probability rule is used for SSVS. We report the model size (MS) and false positives (FP) under the point estimate. This point estimate is also used for prediction under PenCred, Lasso, and elastic net, while the posterior predictive distribution is used under BVS-SL and SSVS. We look at the predictive performance in terms of out of sample mean squared error (RMSPE) and out of sample coverage of 95\% predictive intervals (COV$_{95}$). The coverage refers to the proportion of test sample values contained within predictive intervals. The predictive intervals correspond to credible intervals for the Bayesian approaches BVS-SL and SSVS, whereas for PenCred, as well as the frequentist approaches, they correspond to pseudo confidence intervals that are constructed as $({\bf x}\hat{\bfb}-1.96\sigma_0, {\bf x}\hat{\bfb}+1.96\sigma_0)$, where $\sigma_0$ is the true residual variance. 

It is seen from Tables 1--4 that while the proposed approach has comparable performance in terms of out of sample prediction and 95\% coverage for lower dimensions, the relative performance with respect to competitors improves significantly as the number of noise predictors is increased. The number of true covariates (MS - FP) detected under the proposed approach, as well as the coverage, is essentially always the best or the second best among all the approaches considered. We also see that while the SSVS might have an advantage compared to BVS-SL in terms of controlling false positives, the prior corrected BVS-SL essentially has similar or better multiplicity control compared to SSVS. 


An important observation is that the proposed approach seems to have a distinct advantage over the competitors for a large number of noise variables, a scenario often encountered in high--dimensional genomics applications. In particular, as the number of noise variables is increased: (i) the out of sample coverage under the proposed approach does not change significantly, whereas the coverage for other competitors (except SSVS) decreases; and (ii) the competing approaches (except SSVS) demonstrate the well-known multiplicity problem, registering an increasing number of false positives. On the other hand, the SSVS demonstrates drawbacks in higher dimensions in terms of collinearity, as evidenced by smaller model sizes, and poor power to detect true positives for a given level of false discovery.  

{\noindent \bf Sensitivity to Prior Knowledge:} To evaluate the sensitivity of the prior corrected BVS-SL to prior graphical knowledge, we examined the performance when the prior knowledge deviates from the truth. In particular, we looked at the results when the available prior graph information has $p,2p,3p$ mis-specified edges, and compared it with the scenario when the prior graph was the truth. Our analysis indicate that (i) when the mis-specification error rate is low, the prior corrected BVS-SL has superior performance compared to BVS-SL for high values of the belief parameter, and (ii) when the mis-specification rate is high, the prior corrected BVS-SL can perform poorly for high values of the belief, but performs similarly to unsupervised BVS-SL in uninformative settings involving low values of belief. This is consistent with our expectations in the sense that we can protect against mis-specification through low values of the belief parameter, while hoping to achieve significant improvements through high values of the belief when the available prior graph knowledge  is true or close to the truth. In applications, we expect the belief parameter to be chosen by experts using subject knowledge, or alternatively one can look at a series of beliefs. Figure 4 shows the plot of area for case I(b) under the ROC and PRC curves using our approach incorporating true prior knowledge, and a range of values of the belief parameter. It is clear that under a true prior graph, the area under the curve for prior corrected BVS-SL is a non-decreasing function of the belief parameter.
\begin{figure}
\centering
		\mbox{\includegraphics[height=3in, width= 1 \textwidth]{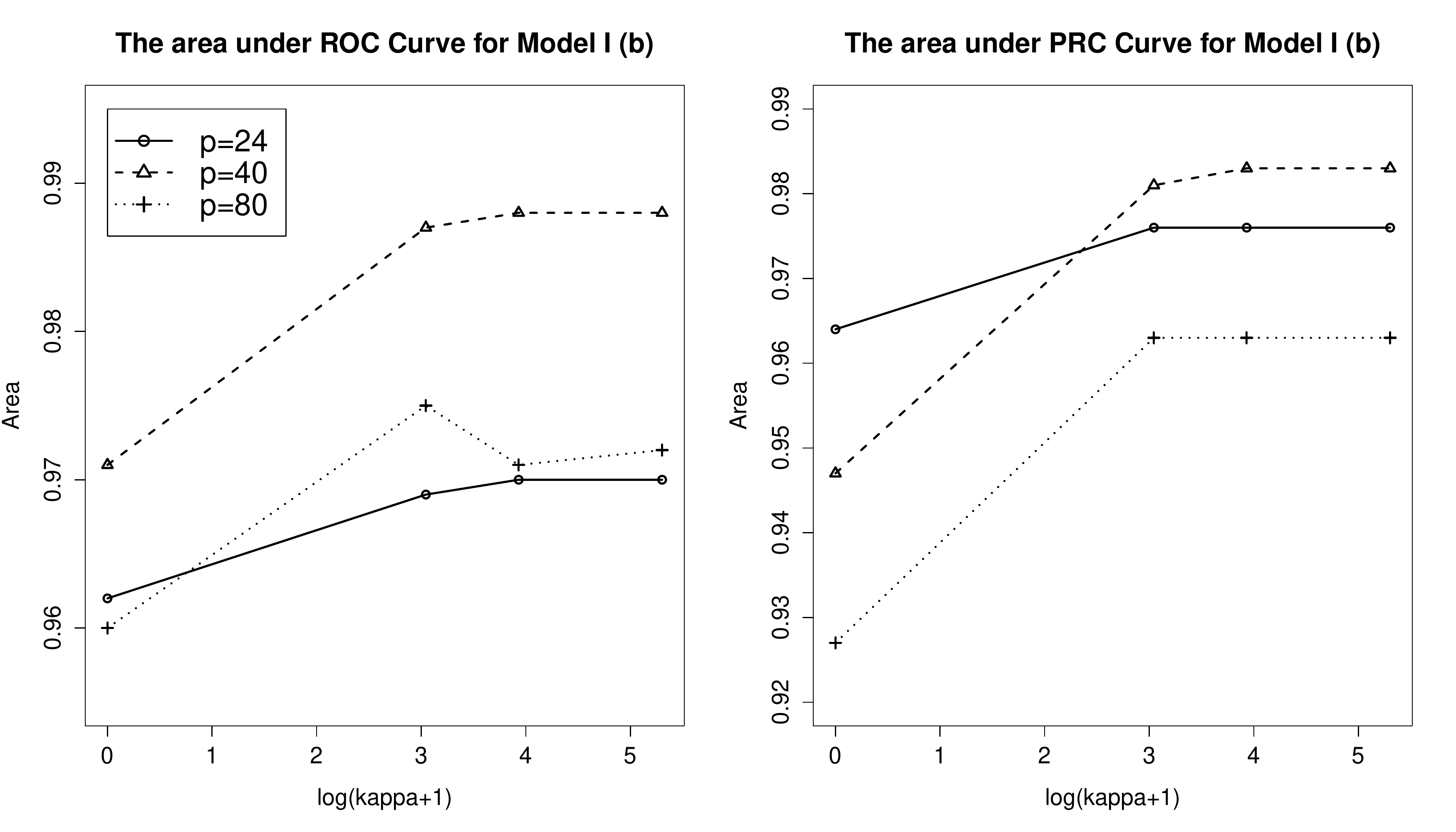}} 		
		\caption{\small{Plots for area under receiver operating characteristic curve and precision recall curve, for varying belief parameters. }}
\end{figure} 

\vskip 12pt

\begin{center} {\bf 6. APPLICATION TO INTEGRATIVE GENOMICS } \end{center} 

{\noindent \bf {6.1. Data Description}}\\
Our motivating dataset arises from a  TCGA-based study in glioblastoma multiforme (GBM), which is the most common and aggressive form of primary brain cancer in human adults (Holland, 2000). The TCGA data portal \begin{small}({http://tcga-data.nci.nih.gov/tcga/})\end{small} provides multiple levels of molecular data for a large cohort  of GBM tumor specimens.  These  specimens underwent rigorous quality control procedures via the TCGA Biospecimen Core to ensure high-quality DNA and RNA extraction. Each qualified specimen was  assayed using multiple assays among which we concentrate on the following: messenger RNA (mRNA) expression using HT-HG-U133A (Affymetrix) arrays, DNA methylation (METH) using HumanMethylation27K (Illumina) and DNA copy number (CN) HG-CGH-244A (Agilent) arrays. All the resulting data from the three platforms are pre-processed, normalized and annotated to the gene level, and referred to as Level 3 data (see Wang et al., 2013 for details). We focus our analysis on 48 genes that overlap with the three critical signaling pathways - RTK/PI3K, p53, and Rb, which are involved in migration, survival and apoptosis progression of cell cycles (Furnari et al., 2007).  Thus our covariate matrix consists of $48$  genes mapped to these core pathways from $D^* = 3$ platforms (mRNA, METH, CN) resulting in $p=48 \times 3= 144$ regressors. Note that mRNA and METH are continuous, while CN is converted into categorical via thresholding, having three categories corresponding to loss, gain, or neutral. The outcome is log-transformed survival times for 163 uncensored patients which is regressed on the covariates using an accelerated failure time model. An integrative study is required for a better understanding of the interactions between these core pathways, and is expected to provide new insights into the progression of GBM (Verhaak et al., 2010). A description of our pre-processing of the copy number data is provided in the Supplementary Materials S.1.

The prior knowledge on the graphical structure between these 48 genes is based on prior studies in GBM (Cerami et al., 2012), and is denoted as $G_{0,pr}$ (shown in panel (b) of Figure 1). This prior graph is obtained by assessing sequence mutations, copy number alterations and proteins and confirm and extend the observation that GBM alterations tend to occur within specific functional modules. Note that while the prior graph has only 48 nodes, the graph in our analysis comprises 144 nodes, across the 3 platforms. We construct the prior graph for these 144 nodes as one which is designed to preserve the prior graphical structure $G_{0,pr}$ within the platforms, but specify null interactions between two distinct genes across two different platforms, while assuming the presence of an edge between different platforms at the same gene location. Thus the prior graph can be concisely written as: $G_0 =  G_{0,pr} \otimes I_3$, where $\otimes$ represents the Kronecker product of the two matrices. 


{\noindent {\bf 6.2. Analysis Results}}\\
{\noindent {\bf  Survival Association:}} The marginal inclusion probabilities under our analysis are presented in Table 0.1 in Supplementary Material, with a corresponding plot in Figure 5. Since a large number of features have inclusion probabilities greater than $0.5$, we select the important features using a false discovery rate criteria controlled at a pre-specified level $\alpha$, as detailed in Appendix A.4. Under a false discovery rate threshold of $\alpha=0.2$ (corresponding to a posterior probability threshold of 0.7), seven genes are significantly associated with progression through various mechanisms: through copy number, we have HRAS, TP53, CCND1, BRAF and CDKN2C; through methylation, we have GRB2  and through mRNA, MDM2. Of these CDKN2C  and GRB2 are positive drivers of progression, while the remaining genes are negatively associated with progression. HRAS is a member of the RAS oncogene family, whose negative effect on Glioblastoma is previously observed on the overall and progression-free survival (Serao et al., 2011). Cyclin D1 (CCND1) belongs to the Cyclin D family of cell cycle regulators, which are known to be up-regulated and amplified in malignant glioma (Buschges et al., 1999). Similarly, MDM2 the inhibitor of the tumor suppressor TP53, is established to be a candidate gene associated with short progression (Lukashchuk and Vousden, 2007). TP53 copy-number itself is associated with poor progression. The effect is through the deletion of TP53, which is known be associated poor progression in GBM (Yin et al., 2009). Although, there is no evidence of BRAF amplification in GBM, a previous study established that BRAF amplification via gene duplication event activates the MAPK signaling in low-grade glioma (Pfister et al., 2008). Cyclin-dependent kinase 4 inhibitor C (CDKN2C) is a well characterized tumor suppressor gene associated with many cancers and known to be deleted in Glioblastoma (Solomon et al., 2008; Dunn et al., 2012).  On the other hand, Growth factor receptor-bound protein 2 (GRB2) is a key protein in epidermal growth factor receptor (EGFR and EGFRVIII) signaling in the Glioblastoma tumoroginesis pathway (Huang et al., 2009). The positive association of progression with DNA methylation of this gene is interesting, as it essentially inhibits the expression of this key regulator, GRB2, in GBM.  
\begin{figure}
\centering
		\mbox{\includegraphics[height=3in, width= 1 \textwidth]{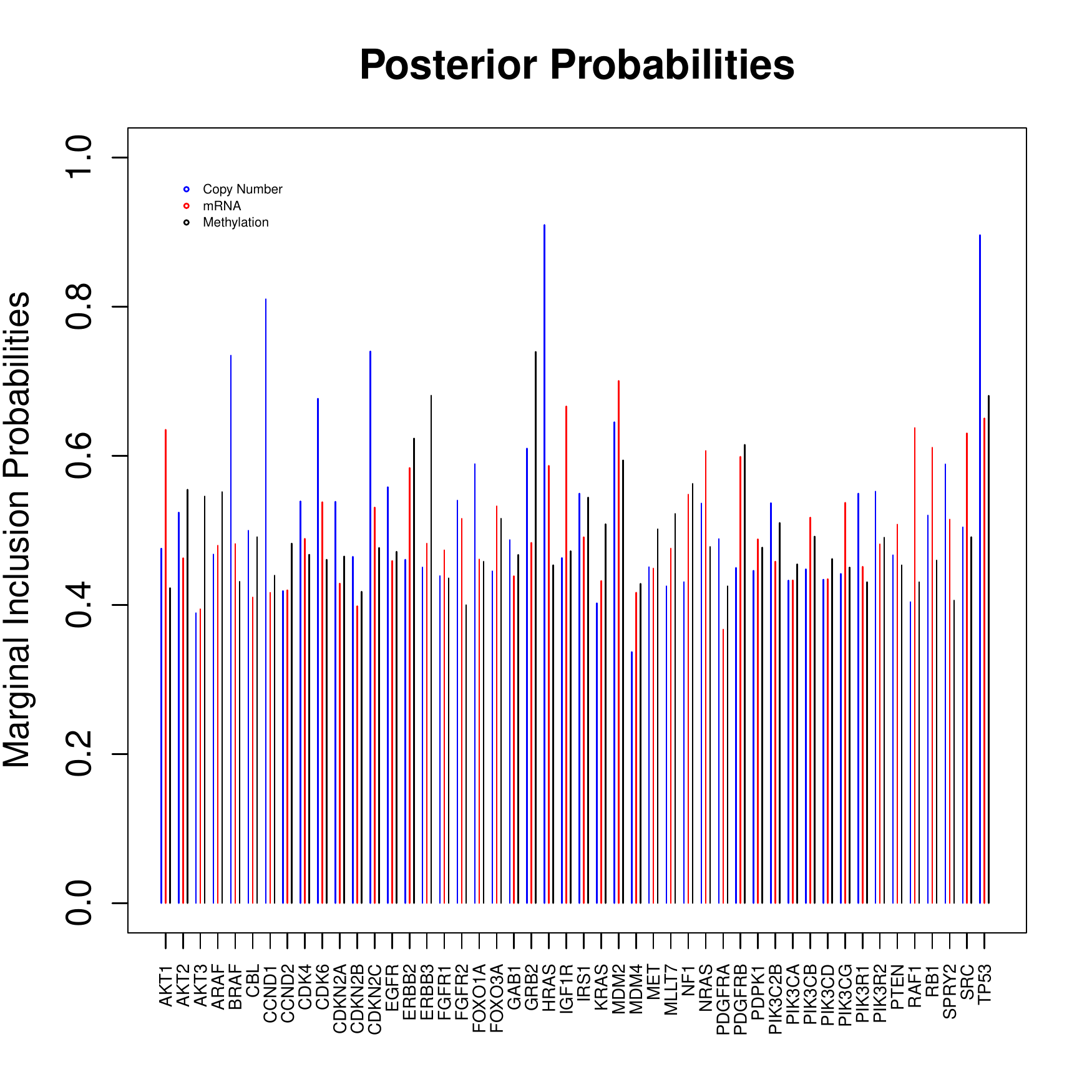}} 		
		\caption{\small{Marginal inclusion probabilities for each gene over three platforms. The probabilities are presented for the three platforms grouped by genes, with blue, red and black, implying copy number, mRNA expression, and methylation, respectively. }}
\end{figure} 

{\noindent \bf Clique Analysis:} The cliques represent groups of associated features within and across platforms which influence the outcome in a coordinated manner. The important cliques are detected as those having highly significant marginal inclusion probabilities $P(\gamma_{C_j}|-), j=1,\ldots,q$. The clique analysis depicted multiple interesting interactions, although all the significant cliques comprise only two-way interactions. In certain cases, the multiple cliques formed with the same molecular probe with different partners have highly significant marginal inclusion probabilities. For instance AKT1 (METH) clique interaction with many different molecular probes is significant (refer Table 0.2 of Supplementary Material).  These cliques constitute both tumor suppressing as well as activating interactions. The cliques involving AKT1 (METH), PTEN (mRNA) and AKT1 (METH), PIK3R2 (mRNA) can be construed as tumor suppressing, while cliques involving AKT1 (METH), CCND1 (mRNA) and AKT1 (METH), GRB2 (CN) probably are tumor activating. The diverse biological functionality of the cliques represent the inherent biological subtypes with in the GBM (Verhaak et al., 2010).

{\noindent \bf Neighborhood Analysis:} In addition to detecting important markers for GBM, we also examine the estimated graph (panel (c) of Figure 1) within and across platforms. The degree of connectivity for nodes with marginal inclusion probabilities $>0.5$ is shown in Table 0.1 in Supplementary Materials. To further explore the biological ramifications of the connectivity, the resulting neighborhood of GRB2 (mRNA) was analyzed using the IPA \begin{small}(Ingenuity Systems, www.ingenuity.com)\end{small} to identify the biological interactions based on literature (Figure 6). A strong effect of DNA methylation is observed in the neighborhood along with the mRNA, while the effect of copy number alteration is more subtle. GRB2 plays a central biological role in this molecular network as a trigger of the RAS signaling upon the activation of upstream receptor tyrosine kinase family members. The presence of three important tumor suppressor genes of GBM: RB1, CDKN2B and PIK3CG is interesting, although they have no direct interaction with GRB2. RB1 and PIK3CG seem to lose their functionality through DNA methylation, while CDKN2B through copy number loss, enabling the RTK/RAS activation cascade via GRB2. These events reinforce the previous illustration in GBM that hypermethylation and deletion of RB1 and CDKN2B respectively contribute to the loss of tumor suppressor function (Nakamura et al., 2001; Rao et al, 2010). 
\begin{figure}
\centering
		\mbox{\includegraphics[height=3in, width= 1 \textwidth]{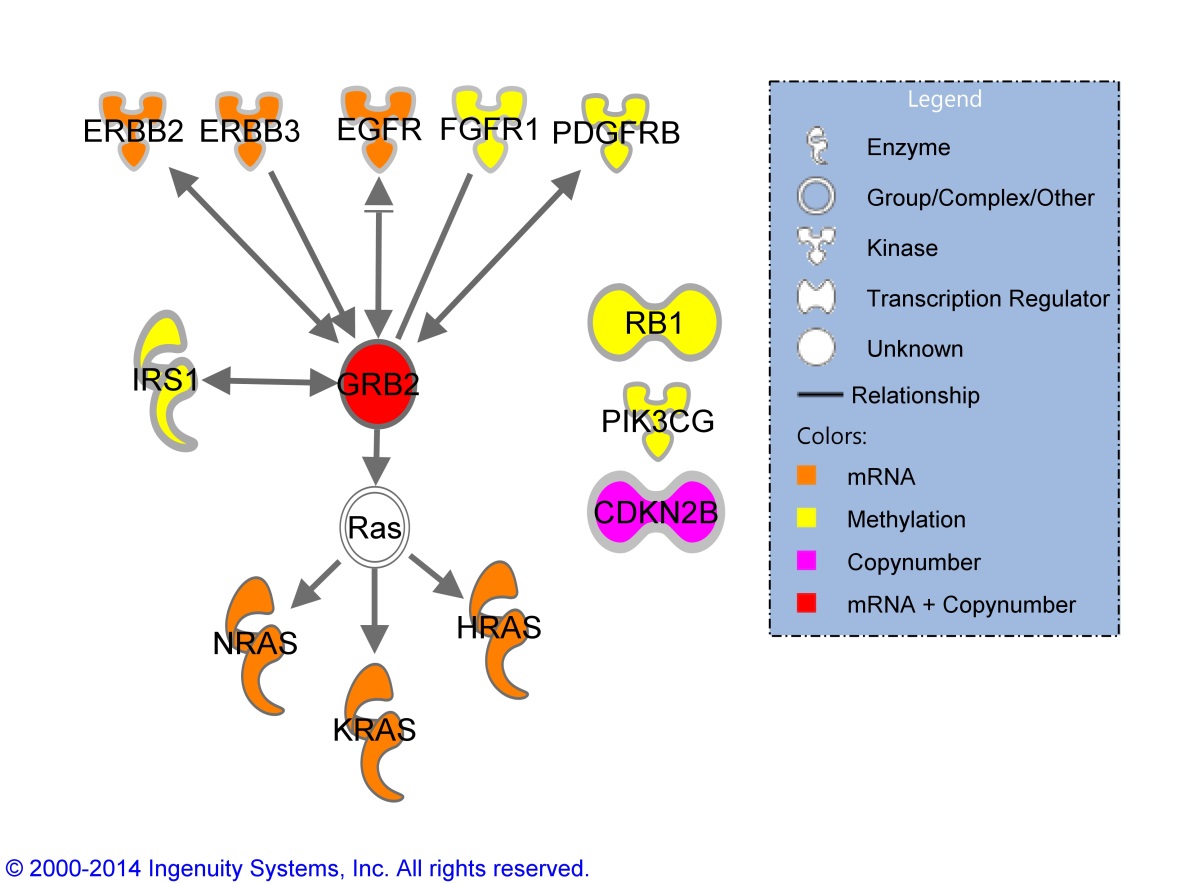}} 		
		\caption{\small{Neighborhood analysis of GRB2 (mRNA expression) using Ingenuity Pathway Analysis,
showing the central biological role in this molecular network as a trigger of the RAS signaling upon the
activation of upstream receptor tyrosine kinase family members.}}
\end{figure} 

The partial correlations of genes between the platforms was further explored to analyze their clustering patterns so as to illustrate globally coordinated changes across platforms. The clustered partial correlations between the genes across the platforms are visualized using the next-generation clustering heatmaps in Figure 7. From the Figure, it is clear that there is a enrichment of positive correlations between the mRNA and copy number data, and an enrichment of negative correlations between the mRNA and DNA methylation data, which further supports our  biological--hypothesis driven integrative models.
\begin{figure}
\centering
		\mbox{\includegraphics[height=3in, width= 1 \textwidth]{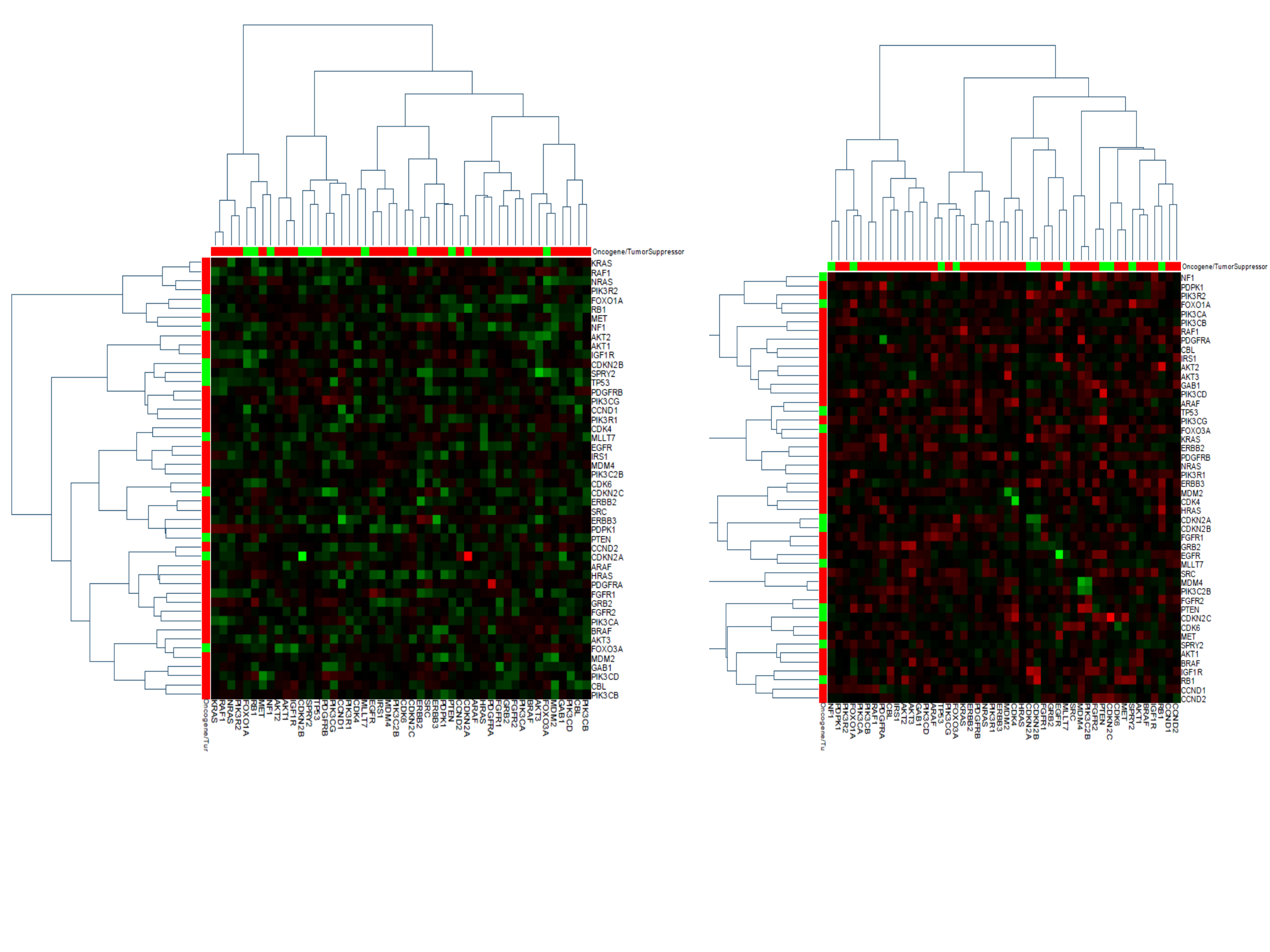} 	}	
		\caption{\small{Left heatmap: hierarchical clustering of correlation between the mRNA and copy number data; right heatmap: hierarchical clustering of correlation between the mRNA and DNA methylation data. Green and red pertains to positive and negative partial correlations respectively.}}
\end{figure} 

\begin{center} {\bf 7. DISCUSSION } \end{center} We have proposed a novel two-step Bayesian structured variable selection approach, which is equipped to learn the graphical structure from mixed data in the presence of prior knowledge, and subsequently uses such structure learning to inform variable selection in a manner that controls for collinearity, while having a desirable control over multiplicity. In this paper, we focussed on integrating (more upstream) copy number, mRNA expression and methylation markers associated with cancer progression; however our methods can be easily extended to account for downstream post-  transcription and translational events such microRNA and proteomics markers. This will provide vital clues towards understanding the complete genomic landscape of cancer development and progression. We leave this task for future consideration.

\vskip 12pt

\begin{center}{ \bf ACKNOWLEDGEMENTS} \end{center}

 V. Baladandayuthapani was partially supported by NIH grant R01 CA160736 and the Cancer Center Support Grant (CCSG) (P30 CA016672). We also thank Hao Wang for providing Matlab code and Lee Ann Chastain for editorial revisions. 

\vskip 12pt
\singlespacing

{\noindent \bf Appendix}

\doublespacing
{\noindent \bf A.1. Regulatory Conditions for Theorem 1: } Let $M_{N}$ be the null model and let $G_0$ now represent the true graph and not prior graphical knowledge as in the paper. Consider the following regulatory conditions:\\
{\noindent \em (A1)} For any model $M_{\bfg'}$ which does not contain the true model \begin{small}$M_{\bfg_0}\ne M_{N}$, $\lim_{n\to\infty} \frac{1}{n}\bfb^T_{\bfg_0}X^T_{\bfg_0}(I - P_{\bfg'})X_{\bfg_0}\bfb_{\bfg_0}=b_{\bfg'}\in(0,\infty)$\end{small}, where $P_{\bfg'}$ is the projection matrix on to the span of ${\bf X}_{\bfg'}$. \\
{\noindent \em (A2)} The true graph $G_0$ has disjoint cliques $\mathcal{C}_1,\ldots,\mathcal{C}_{q_0}$, and the true model $M_{\bfg_0}$ (non-null) admits the representation \begin{small}$\bfg_0 = \cup_{k\in S_0} \{\gamma_{0j}: j\in \mathcal{C}_{k}\}$, for some $S_0\subseteq \{1,\ldots,q_0\}$\end{small}.\\
{\noindent \em (A3)} The graphical model estimation approach is consistent, i.e. \begin{small}$P(\hat{G}=G_0|{\bf X})\to 1$ as $n\to\infty$\end{small}.\\

The first assumption is a standard one in parametric consistency literature for linear models (Liang et al., 2008). The second assumption assumes non-overlapping cliques in $G_0$ and that the true model includes or excludes predictors belonging to a particular clique simultaneously. In the context of genetic studies, disjoint cliques could correspond to distinct clusters of genes or pathways. 
The third assumption regarding graphical model consistency is a strong but necessary condition. In solving the challenging problem of jointly estimating the graphical model and the regression model, there is little hope of acheiving model selection consistency in the second stage in the absence of a consistent graphical model approach in the first stage. We note similar assumptions involving knowledge of the true graph structure have been previously made in frequentist supervised clustering and feature selection literature (Shen et al. 2012). \\

{\noindent \bf A.2. Proof of Theorem 1:}  Let $\mathcal{G}$ denote the finite graph space. First note that 
\begin{small}
\begin{eqnarray*}
P(M_{\bfg_0}|Y^n,{\bf X})&=&\sum_{\hat{G}\in\mathcal{G}} P(M_{\bfg_0}|Y^n,{\bf X},\hat{G})P(\hat{G}|{\bf X}) \\
&=& P(M_{\bfg_0}|Y^n,{\bf X},G_0)P(\hat{G}=G_0|{\bf X}) + \sum_{\hat{G}\in\mathcal{G}\cap \{G_0\}^c} P(M_{\bfg_0}|Y^n,{\bf X},\hat{G})P(\hat{G}|{\bf X}).
\end{eqnarray*}
\end{small}
For a consistent graphical model estimation approach such that $P(\hat{G}=G_0|{\bf X})\to 1$, we only need to show that $P(M_{\bfg_0}|Y^n,{\bf X},G_0)\to 1$ as $n\to\infty$ for model selection consistency. Similarly, it is clear that the predicted value under assumption ({\em A3}) can be expressed as 
\begin{small}
\begin{eqnarray*}
\hat{y}^* &=& \hat{\alpha} + \sum_{\bfg\in \Gamma}\sum_{G\in\mathcal{G}}P(G|{\bf X}){\bf x}^*_{\bfg}\hat{\beta}_{\bfg}P(M_{\bfg}|Y^n,{\bf X},G)\int_{0}^\infty \frac{g}{1+g}\pi(g|M_{\bfg},Y^n,{\bf X}) dg \\
&\approx& \hat{\alpha} + \sum_{\bfg\in \Gamma}{\bf x}^*_{\bfg}\hat{\beta}_{\bfg}P(M_{\bfg}|Y^n,{\bf X},G_0)\int_{0}^\infty \frac{g}{1+g}\pi(g|M_{\bfg},Y^n,{\bf X}) dg \mbox{ for large } n.
\end{eqnarray*}
\end{small}
Let $\mathcal{C}_1,\ldots,\mathcal{C}_{q0}$ be the disjoint cliques corresponding to the true graph $G_0$. Under our approach, the posterior probabilities \begin{small}$P(M_{\bfg}|Y^n,{\bf X})\approx P(M_{\bfg}|Y^n,{\bf X},G_0)$ as $n\to\infty$\end{small}, will only be positive for models belonging to \begin{small}$\Gamma^*=\{\bfg:\bfg=\cup_{k\in S^*} \{\gamma_{0j}: j\in \mathcal{C}_{k}, \forall S^*\subseteq \{1,\ldots,q_0\}\}$\end{small}, so that the true model $M_{\bfg_0}$ defined in (\ref{eq:TrueModel}) and assumption ({\em A2}) belongs to $\Gamma^*$. The proof of the variable selection consistency follows by Theorem 3 of Liang et. al (2008) by using clique indicators $\gamma_{\mathcal{C}_j},j=1,\ldots,q_0$ in place of variable inclusion indicators in their article. The proof for prediction consistency follows using variable selection consistency and using Theorem 4 of Liang et. al (2008). \\

{\noindent \bf A.3. Posterior Computation Steps for Variable Selection} \\
The computation strategy described in section 3 yields an estimate of the graph, which is used to inform the variable selection approach in the second step as described here. At a particular MCMC iteration, let \begin{small}$\bfg^+(j)$\end{small} denote the vector of current variable inclusion indicators having $\gamma_k=1$ for all $k\in C_j$, and similarly denote \begin{small}$\bfg^-(j)$\end{small} as the vector of current variable inclusion indicators having $\gamma_k=0$ for all $k\in C_j$. Further let  \begin{small}$p_{j+}, p_{j-}$\end{small} denote the model sizes corresponding to \begin{small}$\bfg^+_j,\bfg^-_j$\end{small} respectively. The Gibbs sampling alternates as follows\\
\begin{small}
{\noindent Step 1:} Sample $\gamma_{C_j}\sim Ber(\frac{p^+_j}{p^+_j + p^-_j}),j=1,\ldots,p$, where 
\begin{small}
\begin{eqnarray*}
 p^+_{j} &=& \frac{a-2}{p_{j+}+a-2}{}_{2}F_{1}\bigg(\frac{n-1}{2},1;\frac{p_{j+}+a}{2};{\bf R}^2_{\bfg^+_j} \bigg), \quad 
 p^-_{j} = \frac{a-2}{p_{j-}+a-2}{}_{2}F_{1}\bigg(\frac{n-1}{2},1;\frac{p_{j-}+a}{2};{\bf R}^2_{\bfg^-_j} \bigg), \\
 &\mbox{ with }& {}_{2}F_{1}(a,b;c;z) = \frac{\Gamma(c)}{\Gamma(b)\Gamma(c-b)}\int_0^1 \frac{t^{b-1}(1-t)^{c-b-1}}{(1-tz)^a}dt,
\end{eqnarray*}
\end{small}
with ${}_{2}F_{1}(\cdot)$ being the Gaussian hypergeometric function (computed using the `BAS' package in R), ${\bf R}^2_{\bfg}$ is the ordinary coefficient of determination of regression model, and $q$ is the number of cliques. \\
{\noindent Step 2:} Sample fixed effects using 
\begin{small}
$\pi(\bfb_{\bfg}|-) = N \left(\frac{g}{1+g}({\bf X}_{\bm \gamma}^T {\bf X}_{\bm \gamma})^{-1}{\bf X}_{\bm \gamma}^T Y^n, 
\frac{g}{1+g} \frac{({\bf X}_{\bm \gamma}^T {\bf X}_{\bm \gamma})^{-1}}{\eta}\right)$. \end{small}\\
{\noindent Step 3:} Sample the residual precision using 
\begin{small}
$\pi(\eta|-) = Ga(n/2+a_\eta,\sum\limits_{i} (Y^n- {\bf X}_{\bm \gamma^{+} }{\bm \beta}_{\bm \gamma^{+} })^T
 (Y^n- {\bf X}_{\bm \gamma^{+} }{\bm \beta}_{\bm \gamma^{+} })/2+b_{\eta})$.\end{small} \\
{\noindent Step 4:} Sample clique prior inclusion probabilities using \begin{small}$
f(\pi|-) = Beta(q^*+a_{\pi},q-q^*+b_{\pi})$ \end{small}, where $q^*$ is the number of cliques selected using Step 1. \\
{\noindent Step 5:} Sample $g$ using the fact that a hyper-$g$ prior on $g$ with $a=4$ is equivalent to a uniform prior 
on $g'$, where $g' = \frac{g}{1+g}$. For sampling $g'$, we use a discretized grid of $1000$ grids point from $0.01$ to $0.999$ with the posterior probability for grid-point $g^*$ being computed as \begin{small} 
$\pi(g'|-) \propto  \left( \frac{g'}{1-g'} \right)^{-\frac{p_{\bm \gamma}}{2}} \exp\left\{
-\frac{\eta}{2g'/(1-g')} {\bm \beta}_{\bm \gamma}^T ({\bf X}_{\bm \gamma}^T{\bf X}_{\bm \gamma}) {\bm \beta}_{\bm \gamma} 
\right\}. $\end{small} We then use the transformation $g = \frac{g'}{1-g'}$ to obtain $g$. \end{small}\\

{\noindent \bf A.4. FDR based Criteria for Inclusion:} We briefly discuss the FDR based criteria used in section 6.2 to determine the cut-off on posterior probabilities, as described in Baladandayuthapani et al. (2010). In particular, we can choose a threshold $\phi_\alpha$ for posterior probabilities so as to control the average Bayesian FDR at level $\alpha$, which essentially implies that we expect $100\alpha\%$ of the significant markers to be false positives. To obtain such an estimate, first sort the posterior probabilities for all markers in ascending order to yield \begin{small}$pr_{(j)},j=1,\ldots,p$. Then $\phi_\alpha=pr_{(\zeta)}$, where $\zeta=\max\left\{j^*: j^{*-1}\sum_{j=1}^{j^*}pr_{(\zeta)}\le \alpha\right\}$.\end{small}

\begin{small}
\begin{center}
\begin{table}
\caption{Simulations for Case I(a), training sample size = 100, test sample size = 100. }
\begin{tabular}{|l|c|c|c|c|c|c|c|} 
\hline 
\multicolumn{8}{|c|}{p=24}\\
\hline
Method    &  MSPE  & ROC & PRC &Pwr({\footnotesize{10\% FDR}}) & MS & FP & $Cov_{95}$ \\
\hline
BVS-SL & 1.102 & 0.998 & 0.999 & {\bf 1.000} & 10.368 & 0.474 & 0.923 \\ 
Prior Corr BVS-SL ($\kappa=20$) & {\bf 1.101} & {\bf 1.000} & {\bf 1.000} & {\bf 1.000} & 10.250 & {\bf 0.250} &  0.923 \\ 
  PenCred & 1.122 & 0.915 & 0.926 & 0.833 & 8.200 & 0.450 & 0.914 \\ 
  SSVS & 1.107 & 0.960 & 0.957 & 0.895 & 8.200 & 0.300 & 0.{\bf 924} \\ 
  Lasso & 1.157 & 0.892 & 0.907 & 0.815 & 11.400 & 2.700 & 0.912 \\ 
  EL & 1.165 & 0.899 & 0.911 & 0.820 & 11.600 & 2.850 & 0.910 \\ 
\hline
\multicolumn{8}{|c|}{p=40} \\
\hline
BVS-SL & 1.103 & 0.997 & 0.996 & {\bf 1.000} & 10.000 & 0.444 & {\bf 0.922} \\ 
Prior Corr BVS-SL ($\kappa=20$) & {\bf 1.100} & {\bf 1.000} & {\bf 1.000} & {\bf 1.000} & 10.250 & 0.250 & 0.921 \\ 
  PenCred & 1.153 & 0.890 & 0.869 & 0.802 & 9.400 & 1.700 & 0.912 \\ 
  SSVS & 1.123 & 0.954 & 0.921 & 0.880 & 7.650 & {\bf 0.200} & 0.917 \\ 
  Lasso & 1.206 & 0.894 & 0.870 & 0.815 & 10.950 & 2.900 & 0.893 \\ 
  EL & 1.218 & 0.906 & 0.874 & 0.815 & 11.150 & 3.050 & 0.893 \\ 
\hline
\multicolumn{8}{|c|}{p=80}\\
\hline
 BVS-SL & 1.093 & 0.999 & 0.996 & {\bf 1.000} & 9.800 & {\bf 0.250} & 0.928 \\ 
Prior Corr BVS-SL ($\kappa=20$) & {\bf 1.092} & {\bf 1.000} & {\bf 1.000} & {\bf 1.000} & 10.600 & 0.600 & {\bf 0.930} \\ 
  PenCred & 1.270 & 0.888 & 0.790 & 0.740 & 13.700 & 6.150 & 0.880 \\ 
  SSVS & 1.128 & 0.965 & 0.897 & 0.895 & 8.050 & 0.750 & 0.926 \\ 
  Lasso & 1.288 & 0.879 & 0.770 & 0.800 & 10.700 & 3.100 & 0.883 \\ 
  EL & 1.300 & 0.890 & 0.773 & 0.790 & 11.050 & 3.450 & 0.879 \\ 
\hline
\end{tabular}
\vskip 2pt
\begin{footnotesize}
\singlespacing
MSPE: out of sample predictive MSE; Pwr({\footnotesize{10\% FDR}}) is sensitivity controlling for 90\% specificity; MS: estimated model size; FP: false positives, and $Cov_{95}$ is coverage under 95\% predictive intervals.
\end{footnotesize}
\end{table}
\end{center}
\end{small}

\begin{small}
\begin{center}
\begin{table}
\caption{Simulations for Case I(b), training sample size = 100, test sample size = 100. }
\begin{tabular}{|l|c|c|c|c|c|c|c|} 
\hline 
\multicolumn{8}{|c|}{p=24}\\
\hline
Method    &  MSPE  & ROC & PRC & Pwr({\footnotesize{10\% FDR}}) & MS & FP & $Cov_{95}$ \\ 
\hline
BVS-SL & 1.055 & 0.962 & 0.964 & 0.950 & 7.850 & 0.150 & 0.939 \\ 
  Prior Corr BVS-SL ($\kappa=20$) & 1.055 & {\bf0.969} & {\bf0.976} & {\bf0.950} & 8.450 & {\bf0.050} & 0.940 \\ 
  PenCred & 1.057 & 0.770 & 0.802 & 0.565 & 5.600 & 0.250 & {\bf0.942} \\ 
  SSVS & {\bf 1.052} & 0.820 & 0.842 & 0.645 & 6.000 & 0.200 & 0.941 \\ 
  Lasso & 1.168 & 0.618 & 0.677 & 0.450 & 7.200 & 2.300 & 0.907 \\ 
  EL & 1.175 & 0.638 & 0.693 & 0.460 & 7.300 & 2.300 & 0.905 \\ 
\hline
\multicolumn{8}{|c|}{p=40} \\
\hline
 BVS-SL & 1.108 & 0.971 & 0.947 & 0.955 & 9.100 & 1.000 & 0.924 \\ 
Prior Corr BVS-SL ($\kappa=20$) & 1.111 & {\bf0.987} & {\bf0.981} & {\bf0.985} & 9.550 & {\bf0.650} & 0.923 \\ 
  PenCred & 1.156 & 0.749 & 0.679 & 0.550 & 6.200 & 1.200 & 0.912 \\ 
  SSVS & {\bf 1.107} & 0.842 & 0.785 & 0.695 & 6.400 & {\bf0.650} & {\bf0.925} \\ 
  Lasso & 1.251 & 0.630 & 0.563 & 0.423 & 8.650 & 4.050 & 0.885 \\ 
  EL & 1.265 & 0.648 & 0.598 & 0.465 & 8.300 & 3.450 & 0.881 \\ 
\hline
\multicolumn{8}{|c|}{p=80}\\
\hline
BVS-SL & 1.089 & 0.960 & 0.927 & 0.950 & 8.200 & 0.500 & 0.922 \\ 
Prior Corr BVS-SL ($\kappa=20$) & {\bf1.082} & {\bf0.975} & {\bf0.963} & {\bf0.965} & 8.600 & {\bf0.400} & {\bf0.927} \\ 
  PenCred & 1.212 & 0.738 & 0.561 & 0.518 & 9.450 & 4.450 & 0.888 \\ 
  SSVS & 1.093 & 0.863 & 0.730 & 0.725 & 6.050 & 0.600 & 0.924 \\ 
  Lasso & 1.290 & 0.609 & 0.473 & 0.450 & 6.100 & 2.200 & 0.861 \\ 
  EL & 1.295 & 0.628 & 0.536 & 0.530 & 7.250 & 2.750 & 0.861 \\ 
\hline
\end{tabular}
\vskip 2pt
\begin{footnotesize}
\singlespacing
MSPE: out of sample predictive MSE; Pwr({\footnotesize{10\% FDR}}) is sensitivity controlling for 90\% specificity; MS: estimated model size; FP: false positives, and $Cov_{95}$ is coverage under 95\% predictive intervals.
\end{footnotesize}
\end{table}
\end{center}
\end{small}

\begin{small}
\begin{center}
\begin{table}
\caption{Simulations for Case I(c), training sample size = 100, test sample size = 100. }
\begin{tabular}{|l|c|c|c|c|c|c|c|} 
\hline 
\multicolumn{8}{|c|}{p=24}\\
\hline
Method    &  MSPE  & ROC & PRC & Pwr({\footnotesize{10\% FDR}}) & MS & FP & $Cov_{95}$ \\ 
\hline
BVS-SL & 1.070 & 0.958 & 0.941 & 0.905 & 7.789 & 0.421 & 0.923 \\ 
Prior Corr BVS-SL ($\kappa=20$) & {\bf1.068} & {\bf0.981} & {\bf0.978} & {\bf0.975} & 7.800 & {\bf0.250} & {\bf0.925} \\ 
  PenCred & 1.090 & 0.856 & 0.897 & 0.750 & 7.450 & 0.400 & 0.922 \\ 
  SSVS & 1.080 & 0.873 & 0.899 & 0.780 & 7.200 & 0.300 & 0.923 \\ 
  Lasso & 1.109 & 0.852 & 0.822 & 0.760 & 10.450 & 2.450 & 0.914 \\ 
  EL & 1.101 & 0.885 & 0.903 & 0.810 & 10.400 & 2.150 & 0.915 \\ 
\hline
\multicolumn{8}{|c|}{p=40}\\
\hline
BVS-SL& 1.071 & 0.969 & 0.929 & 0.965 & 7.941 & 0.529 & {\bf0.936} \\ 
Prior Corr BVS-SL ($\kappa=20$) & {\bf1.066} & {\bf0.986} & {\bf0.978} & {\bf0.980} & 8.000 & {\bf0.100} & {\bf0.936} \\ 
  PenCred & 1.118 & 0.879 & 0.860 & 0.780 & 9.150 & 1.750 & 0.921 \\ 
  SSVS & 1.086 & 0.880 & 0.894 & 0.825 & 7.150 & 0.250 & 0.932 \\ 
  Lasso & 1.142 & 0.857 & 0.787 & 0.808 & 11.550 & 3.550 & 0.918 \\ 
  EL & 1.145 & 0.895 & 0.847 & 0.820 & 11.250 & 2.950 & 0.917 \\ 
\hline
\multicolumn{8}{|c|}{p=80}\\
\hline
BVS-SL  & 1.092 & 0.951 & 0.927 & 0.935 & 8.000 & 0.316 & {\bf0.928} \\ 
Prior Corr BVS-SL ($\kappa=20$) & {\bf1.084} & {\bf0.992} & {\bf0.988} & {\bf0.990} & 7.650 & {\bf0.050} & {\bf0.928} \\ 
  PenCred & 1.253 & 0.867 & 0.770 & 0.735 & 12.650 & 5.750 & 0.889 \\ 
  SSVS & 1.109 & 0.895 & 0.796 & 0.745 & 6.600 & 0.200 & 0.921 \\ 
  Lasso & 1.179 & 0.814 & 0.729 & 0.695 & 10.650 & 3.800 & 0.909 \\ 
  EL & 1.189 & 0.861 & 0.770 & 0.738 & 9.900 & 3.100 & 0.910 \\ 
\hline
\end{tabular}
\vskip 2pt
\begin{footnotesize}
\singlespacing
MSPE: out of sample predictive MSE; Pwr({\footnotesize{10\% FDR}}) is sensitivity controlling for 90\% specificity; MS: estimated model size; FP: false positives, and $Cov_{95}$ is coverage under 95\% predictive intervals.
\end{footnotesize}
\end{table}
\end{center}
\end{small}

\begin{small}
\begin{center}
\begin{table}
\caption{Simulations for Case I(d), training sample size = 100, test sample size = 100. }
\begin{tabular}{|l|c|c|c|c|c|c|c|} 
\hline 
\multicolumn{8}{|c|}{p=24}\\
\hline
Method    &  MSPE  & ROC & PRC & Pwr({\footnotesize{10\% FDR}}) & MS & FP & $Cov_{95}$ \\ 
\hline
BVS-SL & 1.068 & {\bf0.839} & {\bf0.846} & 0.675 & 6.650 & 0.500 & 0.930 \\ 
Prior Corr BVS-SL ($\kappa=20$) & 1.066 & 0.823 & 0.845 & {\bf0.705} & 6.350 & {\bf0.450} & 0.931 \\   
  PenCred & 1.086 & 0.745 & 0.770 & 0.557 & 5.800 & 0.850 & 0.923 \\ 
  SSVS & {\bf1.065} & 0.810 & 0.810 & 0.560 & 5.800 & 0.500 & {\bf0.932} \\ 
  Lasso & 1.159 & 0.551 & 0.606 & 0.348 & 6.450 & 2.650 & 0.906 \\ 
  EL & 1.167 & 0.576 & 0.633 & 0.353 & 5.700 & 1.950 & 0.906 \\ 
\hline
\multicolumn{8}{|c|}{p=40} \\
\hline
  BVS-SL & 1.072 & {\bf0.922} & 0.851 & {\bf0.845} & 6.850 & 0.850 & 0.934 \\ 
Prior Corr BVS-SL ($\kappa=20$)  & {\bf1.065} & 0.893 & {\bf0.857} & 0.825 & 6.250 & {\bf0.400} & {\bf0.935} \\ 
  PenCred & 1.126 & 0.691 & 0.622 & 0.480 & 6.450 & 1.700 & 0.915 \\ 
  SSVS & 1.071 & 0.855 & 0.767 & 0.645 & 5.500 & 0.500 & 0.933 \\ 
  Lasso & 1.218 & 0.556 & 0.475 & 0.340 & 5.000 & 1.750 & 0.891 \\ 
  EL & 1.225 & 0.594 & 0.516 & 0.390 & 5.100 & 1.550 & 0.893 \\ 
\hline
\multicolumn{8}{|c|}{p=80}\\
\hline
BVS-SL & 1.109 & {\bf0.937} & 0.788 & {\bf0.865} & 5.900 & 0.900 & {\bf0.919} \\ 
 Prior Corr BVS-SL ($\kappa=20$) & {\bf1.105} & 0.894 & {\bf0.798} & 0.815 & 5.800 & {\bf0.400} & 0.918 \\ 
  PenCred & 1.226 & 0.633 & 0.419 & 0.358 & 7.850 & 4.150 & 0.897 \\ 
  SSVS & 1.124 & 0.840 & 0.655 & 0.628 & 4.600 & {\bf0.400} & 0.914 \\ 
  Lasso & 1.239 & 0.509 & 0.354 & 0.320 & 4.650 & 1.700 & 0.886 \\ 
  EL & 1.248 & 0.596 & 0.406 & 0.355 & 4.800 & 1.700 & 0.881 \\ 
\hline
\end{tabular}
\vskip 2pt
\begin{footnotesize}
\singlespacing
MSPE: out of sample predictive MSE; Pwr({\footnotesize{10\% FDR}}) is sensitivity controlling for 90\% specificity; MS: estimated model size; FP: false positives; $Cov_{95}$: coverage under 95\% predictive intervals.
\end{footnotesize}
\end{table}
\end{center}
\end{small}


\begin{thebibliography}{}

\bibitem{} Albert,  J. H., and Chib, S. (1993),  ``Bayesian Analysis of Binary and Polychotomous Response Data," {\it Journal of the American Statistical Association}, 88, 669--679.

\bibitem{} Bachman, K.E.,  Argani, P., Samuels, Y., Silliman, N., Ptak, J., Szabo, S., Konishi, H., Karakas, B., Blair, B.G., Lin, C., Peters, B.A., Velculescu, V.E., and Park, B.H. (2004), ``The PIK3CA Gene is Mutated with High Frequency in Human Breast Cancers," {\it Cancer Biology \& Therapy}, 3(8), 772--775. 
\bibitem{} Baladandayuthapani, V., Ji. Y., Talluri, R., Nieto-Barajas, L. E., and Morris J. S. (2010), ``Bayesian Random Segmentation Models to Identify Shared Copy Number Aberrations for Array CGH Data," {\it Journal of the American Statistical Association}, 105(492),  1358--1375.

\bibitem{} Baladandayuthapani, V., Talluri, R., Ji, Y., Coombes, K.R., Lu, Y., Hennessy, B.T., Davies, M.A., and Mallick, B.K. (2014), ``Bayesian Sparse Graphical Models for Classification with Application to Protein Expression Data," {\it Annals of Applied Statistics}, In press. http://imstat.org/aoas/next\_issue.html

\bibitem{} Boehm, J. S., and Hahn, W. C. (2011), ``Towards Systematic Functional Characterization of Cancer Genomes," {\it Nature Reviews Genetics}, 12, 487--488.

\bibitem{} Bondell, H. D., and Reich, B. J. (2012), ``Consistent High-dimensional Bayesian Variable Selection via Penalized Credible Regions. {\it Journal of the American Statistical Association}

\bibitem{} Buschges, R., Weber, R.G., Actor, B., Lichter, P., Collins, V.P., and Reifenberger, G. (1999), ``Amplification and Expression of Cyclin D Genes (CCND1, CCND2 and CCND3) in Human Malignant Gliomas," {\it Brain Pathology}, 9(3), 435--442.

\bibitem{} Campbell, I.G., Russell, S.E., Choong, D.Y., Montgomery, K.G., Ciavarella, M.L., Hooi, C.S., Cristiano, B.E., Pearson, R.B., and Phillips, W.A. (2004), ``Mutation of the PIK3CA Gene in Ovarian and Breast Cancer," {it Cancer Research}, 64, 7678--7681.

\bibitem{} Canale, A., and Dunson, D.B. (2011), ``Bayesian Kernel Mixture for Counts," {\it Journal of the American Statistical Association}, 106(496), 1528--1539. 

\bibitem{} Capon, D.J., Seeburg, P.H., McGrath, J.P., Hayflick, J.S., Edman U., Levinson, A.D., and Goeddel, D.V. (1983), ``Activation of Ki-ras2 Gene in Human Colon and Lung Carcinomas by Two Different Point Mutationsm" {\it Nature}, 304, 507--513.

\bibitem{} Cerami, E., Demir, E., Schultz, N., Taylor, B.S., and Sander, C. (2010), ``Automated Network Analysis Identifies Core Pathways in Glioblastoma," {\it PLoS One}, 5(2), e8918. 

\bibitem{} Chin, L., Hahn, W.C., Getz, G., and Meyerson, M. (2011), ``Making Sense of Cancer Genomic Data. {\it Genes and Development}, 25, 534-555.

\bibitem{} Davies, H., Bignell, G.R., Cox, C., Stephens, P., Edkins, S., Clegg, S., Teague, J., Woffendin, H., Garnett, M.J., Bottomley, W., Davis, N., Dicks, E., Ewing, R., Floyd, Y., Gray, K., Hall, S., Hawes, R., Hughes, J., Kosmidou, V., Menzies, A., Mould, C., Parker, A., Stevens, C., Watt, S., Hooper, S., Wilson, R., Jayatilake, H., Gusterson, B.A., Cooper, C., Shipley, J., Hargrave, D., Pritchard-Jones, K., Maitland, N., Chenevix-Trench, G., Riggins, G.J., Bigner, D.D., Palmieri, G., Cossu, A., Flanagan, A., Nicholson, A., Ho, J.W., Leung, S.Y., Yuen, S.T., Weber, B.L., Seigler, H.F., Darrow, T.L., Paterson, H., Marais, R., Marshall, C.J., Wooster, R., Stratton, M.R., and Futreal, P.A. (2002), ``Mutations of the BRAF Gene in Human Cancer," {\it Nature}, 417, 948--954. 

\bibitem{} Davis, J., and Goadrich, M. (2006), ``The Relationship between Precision-Recall and ROC Curves,"  {\it Proc. of 23rd International Conference on Machine Learning}, 233--240.

\bibitem{} Dawid, A.P., and Lauritzen, S.L. (1993), ``Hyper Markov Laws in the Statistical Analysis of Decomposable Graphical Models," {\it The Annals of Statistics}, 21, 1272--1317.

\bibitem{} Dunn, G.P., Rinne, M.L., Wykosky, J., Genovese, G., Quayle, S.N., Dunn, I.F., Agarwalla, P.K., Chheda, M.G., Campos, B., Wang, A., Brennan, C., Ligon, K.L., Furnari, F., Cavenee, W.K., Depinho, R.A., Chin, L., and Hahn, W.C. (2012), ``Emerging Insights into the Molecular and Cellular Basis of Glioblastoma," {\it Genes Development}, 26(8), 756--84.

\bibitem{} Furnari, F. B., Fenton, T., Bachoo, R. M., Mukasa, A., Stommel, J. M., Stegh, A., Hahn, W. C., Ligon, K. L., Louis, D. N., Brennan, C., Chin, L., DePinho, R. A., and Cavenee, W. K. (2007), ``Malignant Astrocytic Glioma: Genetics, Biology, and Paths to Treatment," {\it Genes and Development}, 21, 2683--2710.

\bibitem{} George, E.I., and McCulloch, R.E. (1993), ``Variable Selection via Gibbs Sampling," {\it Journal of the American Statistics Association}, 88(423), 881--889.

\bibitem{} Hamid, J.S., Hu, P., Roslin, N.M., Ling, V., Greenwood, C., and Beyene, J. (2009), ``Data Integration in Genetics and Genomics: Methods and Challenges," {\it Human Genomics and Proteomics: HGP},  1, 1--13.

\bibitem{} Holland, E. C. (2000), ``Glioblastoma Multiforme: The Terminator," {\it Proceedings of the National Academy of Sciences}, 97, 6242--6244.

\bibitem{} Huang, P.H., Xu, A.M., and White, F.M. (2009), ``Oncogenic EGFR Signaling Networks in Glioma," {\it Science Signaling}, 2(87), re6.

\bibitem{} Jara, A., Garcia-Zattera, M., and Lesaffre, E. (2007), ``A Dirichlet Process Mixture Model for the Analysis of Correlated Binary Responses," {\it Computational Statistics \& Data Analysis}, 51, 5402--5415.

\bibitem{} Jennings, E.M., Morris, J.E., Carroll, R.J., Manyam, G.C., and Baladandayuthapani, V. (2013), ``Bayesian Methods for Expression-based Integration of Various Types of Genomics Data," {\it EURASIP Journal on Bioinformatics and Systems Biology}, 1, 1687--4153.  doi: 10.1186/1687-4153-2013-13.

\bibitem{} Johnson, V.E.,  and Albert, A.H. (1999), {\it Ordinal Data Models}, New York: Springer-Verlag.  

\bibitem{} Kottas, A., M\o{u}ller, P., and Quintana, F. (2005), ``Nonparametric Bayesian Modeling for Multivariate Ordinal Data," {\it Journal of Computational and Graphical Statistics}, 14, 610--625.

\bibitem{} Kundu, S., Baladandayuthapani, V., and Mallick, B.K.  (2014), ``Bayes Regularized Graphical Model Estimation in High Dimensions," {\it http://arxiv.org/abs/1308.3915}. 

\bibitem{} Lanckriet, G.R.G., De Bie, T., Cristianini, N.,  Jordan, M.I., and  Noble, W.S. (2004), ``A Statistical Framework for Genomic Data Fusion," {\it Bioinformatics}, 20, 2626--2635.

\bibitem{} Li, F., and Zhang, N. R. (2010), ``Bayesian Variable Selection in Structured High-dimensional Covariate Spaces with Applications in Genomics," {\it Journal of the American Statistical Association}, 105(3), 1978--2002.

\bibitem{} Liang, F., Paulo, R., Molina, G., Clyde, M.A., and Berger, J.O. (2008), ``Mixtures of g-priors for Bayesian Variable Selection," {\it Journal of the American Statistical Association}, 103(481), 410--423.

\bibitem{} Lock, E.F., Hoadley, K.A., Marron, J.S., and Nobel, A.B. (2013), ``Joint and Individual Variation Explained (JIVE) for Integrated Analysis of Multiple Data Types," {\it Annals of Applied Statistics}, 7(1), 523--542.

\bibitem{} Lukashchuk, N., and Vousden, K.H. (2007), ``Ubiquitination and Degradation of Mutant p53," {\it Molecular Cell Biology}, 27(23), 8284--8295.

\bibitem{} Mukherjee, S., and Speed, T.P., (2008), ``Network Inference Using Informative Priors," {\it Proceedings of the National Academy of Sciences, U. S. A.}, 105, 14313--14318.

\bibitem{} Nakamura, M., Yonekawa, Y., Kleihues, P., and Ohgaki, H. (2001), ``Promoter Hypermethylation of the RB1 Gene in Glioblastomas," {\it Laboratory Investigation}, 81, 77--82.

\bibitem{} Pan, W., Benhuai, X., and Xiaotong, S. (2010), ``Incorporating Predictor Network in Penalized Regression with Application to Microarray Data," {\it Biometrics}, 66(2), 474--484. 

\bibitem{} Pfister, S., Janzarik, W.G., Remke, M., Ernst, A., Werft, W., Becker, N., Toedt, G., Wittmann, A., Kratz, C., Olbrich, H., Ahmadi, R., Thieme, B., Joos, S., Radlwimmer, B., Kulozik, A., Pietsch, T., Herold-Mende, C., Gnekow, A., Reifenberger, G., Korshunov, A., Scheurlen, W., Omran, H., and Lichter, P. (2008), ``BRAF Gene Duplication Constitutes a Mechanism of MAPK Pathway Activation in Low-grade Astrocytomas," {\it Journal of Clinical Investigation}, 118, 1739--1748.


\bibitem{} Rao, S.K., Edwards, J., Joshi, A.D., Siu, I.M., and Riggins, G.J. (2010), ``A Survey of Glioblastoma Genomic Amplifications and Deletions," {\it Journal of Neurooncology}, 96(2), 169--179.

\bibitem{} Rockova, V., and Lesaffre E. (2014), ``Incorporating Grouping Information in Bayesian Variable Selection with Applications in Genomics," {\it Bayesian Analysis}, 9, 221--258.

\bibitem{} Samuels, Y., Wang, Z., Bardelli, A., Silliman, N., Ptak, J., Szabo, S., Yan, H., Gazdar, A., Powell, S.M., Riggins, G.J., Willson, J.K., Markowitz, S., Kinzler, K.W., Vogelstein, B., and Velculescu, V.E. (2004), ``High Frequency of Mutations of the PIK3CA Gene in Human Cancers," {\it Science}, 304, 504--508. 
\bibitem{} Scott, J.G., and Berger, J.O. (2010), ``Bayes and Empirical-Bayes Multiplicity Adjustment in the Variable-Selection Problem," {\it The Annals of Statistics}, 38, 2587--2619.

\bibitem{} Serao, N.V., Delfino, K.R., Southey, B.R., Beever, J.E., and Rodriguez-Zas, S.L. (2011), ``Cell Cycle and Aging, Morphogenesis, and Response to Stimuli Genes are Individualized Biomarkers of Glioblastoma Progression and Survival," {\it BMC Medical Genomics}, 4, 49. doi: 10.1186/1755-8794-4-49.

\bibitem{} Shen, R., Wang, S., and Mo, Q. (2013), ``Sparse Integrative Clustering of Multiple Omics Data Sets," {\it The Annals of Applied Statistics}, 7(1), 269-294.

\bibitem{} Shen, X., Huang, H.C., and Pan, W. (2012), ``Simultaneous Supervised Clustering and Feature Selection Over a Graph,"  {\it Biometrika}, 1-16. doi: 10.1093/biomet/ass038

\bibitem{} Shimizu, K., Birnbaum, D., Ruley, M.A., Fasano, O., Suard, Y., Edlund, L., Taparowsky, E., Goldfarb, M., and Wigler, M. (1983), ``Structure of the Ki-ras Gene of the Human Lung Carcinoma Cell Line Calu-1," {\it Nature}, 304, 487--500.

\bibitem{} Solomon, D.A., Kim, J.S., Jean, W., and Waldman, T. (2008), ``Conspirators in a Capital Crime: Co-deletion of p18INK4c and p16INK4a/p14ARF/p15INK4b in Glioblastoma Multiforme," {\it Cancer Research}, 68, 8657--8660.

\bibitem{} Stingo, F. and Vannucci, M. (2011), ``Variable Selection for Discriminant Analysis with Markov Random Field Priors for the Analysis of Microarray Data," {\it Bioinformatics}, 27(4), 485--501. 

\bibitem{} Stingo, F.C., Chen, Y. A., Tadesse, M.G., and Vannucci, M. (2011), ``Incorporating Biological Information into Linear Models: A Bayesian Approach to the Selection of Pathways and Genes," {\it The Annals of Applied Statistics}, 5(3), 1978--2002.

\bibitem{} Talluri, R., Baladandayuthapani, V., and Mallick, B.K. (2014), ``Bayesian Sparse Graphical Models and Their Mixtures," {\it Stat}, 3(1), 109--125. DOI: 10.1002/sta4.49

\bibitem{} TCGA (2008), ``Comprehensive Genomic Characterization Defines Human Glioblastoma Genes and Core Pathways," {\it Nature}, 455, 1061--1068.

\bibitem{} Tibshirani, R., Saunders, M., Rosset, S., Zhu, J. and Knight, K. (2005), ``Sparsity and Smoothness via the Fused Lasso," {\it Journal of the Royal Statistical Society, Series B, Statistical Methodology}, 67(1), 91--108.

\bibitem{} Tyekucheva, S., Marchionni, L., Karchin, R., and Parmigiani, G. (2011), ``Integrating Diverse Genomic Data Using Gene Sets,"  {\it Genome Biology}, 12(10), R105.

\bibitem{} Verhaak, R.G., Hoadley, K.A., Purdom, E., Wang, V., Qi, Y., Wilkerson, M.D., Miller, C.R., Ding, L., Golub, T., Mesirov, J.P., Alexe, G., Lawrence, M., O'Kelly, M., Tamayo, P., Weir, B.A., Gabriel, S., Winckler, W., Gupta, S., Jakkula, L., Feiler, H.S., Hodgson, J.G., James, C.D., Sarkaria, J.N., Brennan, C., Kahn, A., Spellman, P.T., Wilson, R.K., Speed, T.P., Gray, J.W., Meyerson, M., Getz, G., Perou, C.M., and Hayes, D.N.; Cancer Genome Atlas Research Network. (2010), ``Integrated Genomic Analysis Identifies Clinically Relevant Subtypes of Glioblastoma Characterized by Abnormalities in PDGFRA, IDH1, EGFR, and NF1," {\it Cancer Cell}, 17, 98--110.

\bibitem{} Wang, H. (2012), ``Bayesian Graphical Lasso Models and Efficient Posterior Computation," {\it Bayesian Analysis}, 7(2), 771-790.

\bibitem{} Wang, W., Baladandayuthapani, V., Morris, J.S., Broom, B.M., Manyam, G.C., and Do, K.A. (2013), ``iBAG: Integrative Bayesian Analysis of High-dimensional Multiplatform Genomics Data," {\it Bioinformatics}, 29(2), 149--159. 

\bibitem{} Yin, D., Ogawa, S., Kawamata, N., Tunici, P., Finocchiaro, G., Eoli, M., Ruckert, C., Huynh, T., Liu, G., Kato, M., Sanada, M., Jauch, A., Dugas, M., Black, K.L., and Koeffler, H.P.   (2009), ``High-resolution Genomic Copy Number Profiling of Glioblastoma Multiforme by Single Nucleotide Polymorphism DNA Microarray," {\it Molecular Cancer Research}, 7(5), 665--677.

\bibitem{} Zou, H., and Hastie, T. (2005), ``Regularization and Variable Selection via the Elastic Net," {\it Journal of the Royal Statistical Society, Series B, Statistical Methodology}, 67(2), 301--320.

\end{thebibliography}
\end{document}